\documentclass{jfm}

\usepackage{wrapfig} 
\usepackage{graphicx}

\usepackage{floatrow}
\usepackage{caption,subcaption}
\usepackage{natbib}
\usepackage{bm} 
\usepackage{color} 
\usepackage[usenames,dvipsnames,svgnames,table]{xcolor}
\usepackage{amsmath}
\usepackage{pstricks}
\usepackage{pgf,tikz}
\usepackage{pgfplots}
\pgfplotsset{grid style={dashed,gray}}
\pgfplotsset{minor grid style={dashed,red}}
\pgfplotsset{major grid style={dotted,gray!80!black}}
\pgfplotsset{compat=1.14} 
\graphicspath{{Figures/}}
\usepackage[
	linkcolor=blue,
	anchorcolor=blue,
	citecolor=blue]{hyperref}
\usepackage{graphicx}
\usepackage[font=footnotesize,figurewithin=none]{caption}
\usepackage{subcaption}
\usetikzlibrary{arrows.meta}
\newcommand{\sidecaption}[1]
{\raisebox{\abovecaptionskip}{\begin{subfigure}[t]{1.6em}
			\caption[singlelinecheck=off]{}
			\label{#1}
	\end{subfigure}}\ignorespaces}

\ifCUPmtlplainloaded \else
  \checkfont{eurm10}
  \iffontfound
    \IfFileExists{upmath.sty}
      {\typeout{^^JFound AMS Euler Roman fonts on the system,
                   using the 'upmath' package.^^J}%
       \usepackage{upmath}}
      {\typeout{^^JFound AMS Euler Roman fonts on the system, but you
                   dont seem to have the}%
       \typeout{'upmath' package installed. JFM.cls can take advantage
                 of these fonts,^^Jif you use 'upmath' package.^^J}%
      }
  \else
  \fi
\fi

\ifCUPmtlplainloaded \else
  \checkfont{msam10}
  \iffontfound
    \IfFileExists{amssymb.sty}
      {\typeout{^^JFound AMS Symbol fonts on the system, using the
                'amssymb' package.^^J}%
       \usepackage{amssymb}%
       \let\le=\leqslant  \let\leq=\leqslant
       \let\ge=\geqslant  
      }{}
  \fi
\fi

\ifCUPmtlplainloaded \else
  \IfFileExists{amsbsy.sty}
    {\typeout{^^JFound the 'amsbsy' package on the system, using it.^^J}%
     \usepackage{amsbsy}}
    {\providecommand\boldsymbol[1]{\mbox{\boldmath $##1$}}}
\fi

\newsavebox{\astrutbox}
\sbox{\astrutbox}{\rule[-5pt]{0pt}{20pt}}

\title[Cluster-based network model]{Cluster-based network model}
\author[H.~Li, 
        D.~Fernex, R.~Semaan, 
        J.~Tan,
        M.~Morzy\'nski 
        \& B.~R.~Noack]%
{Hao Li$^{1,2}$,
 Daniel Fernex$^3$,
Richard Semaan$^3$, \\
 Jianguo Tan$^1$$^\dagger$, 
 Marek Morzy\'nski$^4$ \and\ 
 Bernd R. Noack$^{5,2}$%
 \thanks{Email address for correspondence: jianguotan@nudt.edu.cn, bernd.noack@hit.edu.cn}}

\affiliation{
$^1$ Science and Technology on Scramjet Laboratory, \\ 
     National University of Defense Technology, 
     Changsha 410073, Hunan Province, People's Republic of China
\\[\affilskip]
$^2$ Hermann-F\"ottinger-Institut, 
     Technische Universit\"at Berlin,\\
     M\"uller-Breslau-Stra{\ss}e 8, D-10623 Berlin, Germany
\\[\affilskip]
$^3$ Institut f\"ur Str\"omungsmechanik, 
     Technische Universit\"at Braunschweig,\\
     Hermann-Blenk-Stra{\ss}e 37, D-38108 Braunschweig, Germany
\\[\affilskip]
$^4$ Chair of Virtual Engineering,
     Pozna\'n University of Technology,\\
     Jana Paw\l{}a II 24 street, 60-965 Pozna\'n, Poland\\[\affilskip]
$^5$ Center for Turbulence Control,
     Harbin Institute of Technology, Shenzhen, 
     Room 312, Building C, University Town, Xili, Shenzhen 518058, 
     People's Republic of China
\\[\affilskip]
 }

\pubyear{...}
\volume{...}
\pagerange{...}
\date{?; revised ?; accepted ?. - To be entered by editorial office}
\begin{document}
\maketitle


\begin{abstract}

\begin{wrapfigure}[20]{r}[0pt]{0.71\textwidth}
    \vspace*{-8mm}
\begin{flushright}
   \includegraphics[width=100mm]{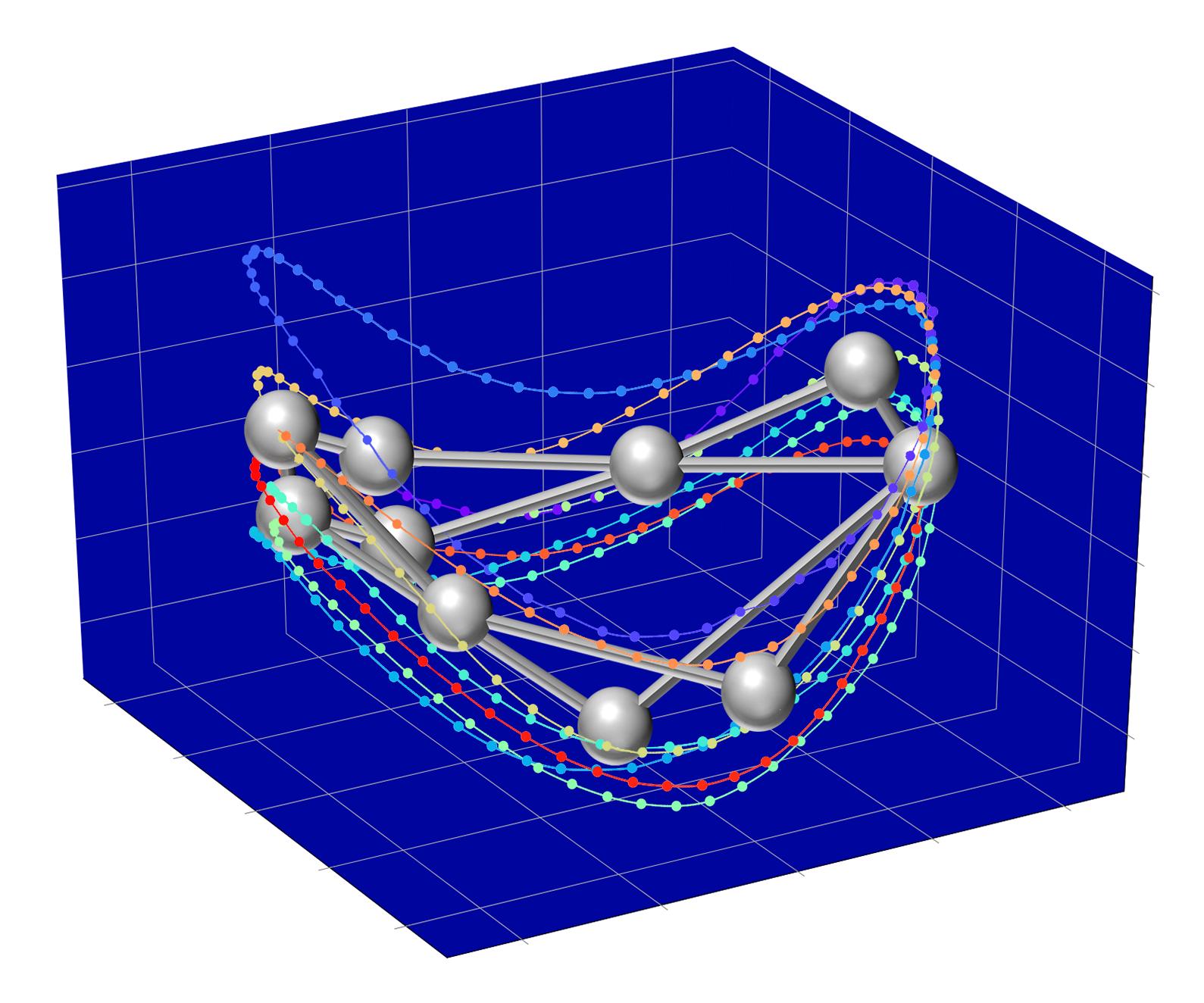}
\end{flushright}
\end{wrapfigure}
We propose an automatable data-driven\break methodology 
for robust nonlinear reduced-order modelling
from time-resolved snapshot data.
In the kinematical coarse-graining,
the snapshots are clustered into few centroids representable for the whole ensemble.
The dynamics is conceptualized as a directed network,
where the centroids represent nodes
and the directed edges denote possible finite-time transitions.
The transition probabilities and times are inferred from the snapshot data.
The resulting cluster-based network model constitutes a deterministic-stochastic
grey-box model resolving the coherent-structure evolution.
This model is motivated by limit-cycle dynamics, 
illustrated for the chaotic Lorenz attractor and 
successfully demonstrated for the laminar two-dimensional mixing layer 
featuring Kelvin-Helmholtz vortices and  vortex pairing,
and for an actuated turbulent boundary layer with complex dynamics.
Cluster-based network modelling opens 
a promising new avenue with unique advantages
over other model-order reductions based on clustering or proper orthogonal decomposition.
\end{abstract}

\section{Introduction}
\label{ToC:Introduction}
We propose a cluster-based network model (CNM)
from time-resolved snapshot data exemplified 
for a laminar mixing layer and an actuated turbulent boundary layer.
The goal is purely data-driven reduced-order modelling
trading the physical insights from first principles,
e.g., the Galerkin method \citep[see, e.g.,][]{Holmes2012book},
with simplicity, robustness and closeness to the original data.

The mixing layer is an archetypical flow configuration 
associated with many academic and industrial applications.
The flow is discussed virtually in any textbook of fluid mechanics.
In the early stage, 
the laminar mixing layer gives rise to periodic, spatially growing Kelvin-Helmholtz vortices
as described in stability theory \citep{Michalke1964jfm},
by  vortex models \citep{Hama1962pf}
or by a Proper Orthogonal Decomposition (POD) Galerkin model \citep{Noack2005jfm}.
At later stages, 
multiple vortex pairings induce 
the inverse cascade to lower wavenumbers and frequencies \citep{Coats1997pecs}.
In addition, three-dimensional instabilities enrich the coherent structures
by rib vortices and spanwise waviness \citep[see, e.g.,][]{Liu1989arfm}.
These mixing layer structures may be seen in the near-field region of wakes and jets.
Moreover,  control of most shear flows, including bluff-body wakes and jets, 
is based on an effective manipulation of the mixing layer
\citep{Fiedler1990pas}.

Another fundamental flow configuration is the turbulent boundary layer.
Since Prandtl's (1904) discovery of the boundary layer theory,
this flow is the cornerstone of practically every fluid and aerodynamic problem. 
In particular, skin-friction reduction  through passive or active means 
has been the subject of research for many decades \citep{Gadelhak2000book,Fan2016book}.
Promising strategies include riblets \citep{walsh_optimization_1984}, 
compliant surfaces \citep{Luhar2016jot}, spanwise wall oscillations 
\citep{Jung1992,Quadrio2009jfm}, and
spanwise traveling  waves  with a Lorentz force \citep{Du2000} 
or wall-normal deflection
\citep{Klumpp2011ftc,Albers2020ftc}.
In this study, 
a spatio-temporal surface deformation 
with  transversal travelling waves is chosen
targeting aerodynamic applications.
Thus, 
a drag reduction of 4.5 \% was experimentally achieved 
for turbulent boundary layer \citep{Li2015}.
In a numerical partner study, 
the actuation parameters were improved 
yielding 31 \% drag reduction \citep{Albers2020ftc, Fernex2020}.
The actuation was also applied over a wing section \citep{Albers2019b}, 
where the pressure varies in the streamwise direction.  
Thus, the total drag was reduced by $7.5\,\%$ accompanied by a slight lift increase.

Since many decades, 
the mixing layer and the turbulent boundary layer 
have been long-standing benchmarks for reduced-order modelling.
For the mixing layer, employed methods include
input-output transfer functions \citep{Sasaki2017jfm},
parabolized stability equations \citep{Sasaki2018tcfd},
vortex filament models \citep{Ashurst1988jfm},
POD models \citep{Delville1999jfm,Ukeiley2001jfm,Wei2009jfm}, and
cluster-based reduced-order models \citep{Kaiser2014jfm}.
Already the laminar two-dimensional shear layer 
can give rise to multiple frequencies \citep{Kasten2016am}.
The early stages of the convectively unstable  and nearly linear dynamics
of mixing layers and jets 
are well resolved by 
parabolized stability equations
requiring little empirical input \citep{Jordan2013arfm}.
After the three-dimensional transition, 
the accuracy of stability-based methods  rapidly deteriorates
or describes only a narrow frequency spectrum of the fluid dynamics.
Stability methods combined with eddy-viscosity closure models 
may significantly extend the application range \citep{Liu1989arfm}.
Alternatively,  data-driven gray-box models from snapshot data 
distilling the coherent-structure dynamics 
become an attractive avenue \citep{Taira2017aiaaj}.

Since the pioneering POD model 
of \citet{Aubry1988jfm}
for the unforced turbulent boundary layer,
numerous advances of data-driven Galerkin models have been proposed.
\citet{Podvin1998jfm}  proposed a low-dimensional model for the minimal channel flow unit 
for the purpose of physical understanding.
Later, 
\citet{Podvin2009pf} has developed an accurate high-dimensional POD model for the wall region of a turbulent channel flow.
The drag-reducing effect of compliant walls has been included in POD  models by  \citet{Lumley1999}.

POD Galerkin methods  arguably constitute the most popular 
and best-investigated data-driven gray-box modelling.
POD Galerkin methods are intimately tied with the Navier-Stokes equations.
While the kinematics, 
the modal expansion, is distilled from data,
the temporal dynamics may be derived from first principles.
Yet, the modal expansion encapsulates a convection dominated dynamics
in an elliptic approach.
This mismatch between the modelling approach and dynamics is the root cause 
of the fragility of data-driven Galerkin models \citep{Noack2016jfm2}.
For the mixing layer, 
the lack of robustness 
is particularly pronounced as exhibited by modelled transients
which are orders of magnitudes too large \citep{Noack2005jfm}.
Moreover, the time integration of the Galerkin model 
may easily lead to states far away from the training data, 
i.e., outside the region of model validity.
This problem persists for other data-driven modal expansions, 
like the dynamic mode decomposition 
\citep{Rowley2009jfm,Schmid2010jfm}.

This robustness challenge of elliptical approaches is avoided
by cluster-based reduced-order models
pioneered by \citet{Burkardt2006siam}.
Here, the state is coarse-grained to a small number of centroids
representative for the whole ensemble of snapshots.
Hence, modelled states will be close 
to the training data by the very construction.
The potential of an extrapolation, 
e.g.\ predicting larger fluctuation amplitudes,
is traded for robustness, 
i.e.,  staying close to the snapshot data.

In the cluster-based Markov model (CMM) for the mixing layer \citep{Kaiser2014jfm,Li2020POF}, 
the temporal evolution is modelled as 
a probabilistic Markov model of the transition dynamics.
The state vector of cluster probabilities 
may initially start in a single centroid 
but eventually diffuses to a fixed point 
representing the post-transient attractor.
This fixed point is well reproduced by CMM.
In addition, CMM has provided valuable physical insights
for the mixing layer and Ahmed body wake \citep{Kaiser2014jfm},
for the turbulent boundary layer \citep{Ishar2019jfm},
for combustion related mixing \citep{Cao2014ef}, 
and for control design \citep{Kaiser2017tcfd,Nair2019jfm}.

A challenge for CMM is the temporal evolution: 
the state may quickly diffuse over the whole attractor,
often within one typical time period. 
This study aims at cluster-based network model (CNM) with improved dynamics resolution
following \citet{Fernex2019pamm}.
The dynamics is modelled by `constant velocity flights' 
between the centroids as `airports'.
The transition probabilities and times 
are consistent with the snapshot data.
The dynamics is thus restricted to a sparse network of routes between the centroids.
Network models are enjoying increasing popularity in
all mathematical modelling fields including biology, sociology, computer sciences.
Network models have also been employed 
to explain vortex dynamics \citep{Nair2015jfm,Taira2016jfm}.
\citet{newman2010networks} provides an excellent introduction to networks.

On the surface, CNM, CMM and POD models 
look like similar data-driven reduced-order models
from snapshot data.
Yet, there are fundamental application differences 
which may be elucidated by an analogy to computational fluid mechanics.
The traditional CMM might be compared 
with unsteady Reynolds Averaged Navier-Stokes (URANS) simulations
converging to the mean flow while resolving some dynamic features during the transient.
In contrast, the proposed CNM mimics a large-eddy simulation 
designed to resolve unsteady coherent-structure dynamics.
The applications of CNM are comparable with POD models. 
The POD model can be conceptualized as a data-driven version 
of the spectral method being routed in the traditional Galerkin methodology.
In contrast, CNM is closer to a collocation method 
using the centroids as `lighthouses' for the corresponding Voronoi cells.

The paper is organized as follows. 
Section \ref{ToC:Method} elaborates the methodology of cluster-based network model (CNM).  
The limit cycle dynamics and Lorenz attractor are employed as illustrating examples. 
Two flow configurations are chosen for the numerical analysis, 
an incompressible laminar mixing layer 
and a turbulent boundary layer.
For the mixing layer (\S~ \ref{ToC:ML}), 
the proposed CNM is benchmarked against the cluster-based Markov model (CMM). 
In \S~ \ref{ToC:TBL}, CNM is performed for 
the three-dimensional actuated turbulent boundary layer featuring a more complex dynamics
\S~\ref{ToC:Conclusions} summarizes this study and outlines future directions of research.

\section{Cluster-based modelling}
\label{ToC:Method}
\begin{figure}
	\centering
	\def\svgwidth{0.8\linewidth}
	 \includegraphics[width=\linewidth]{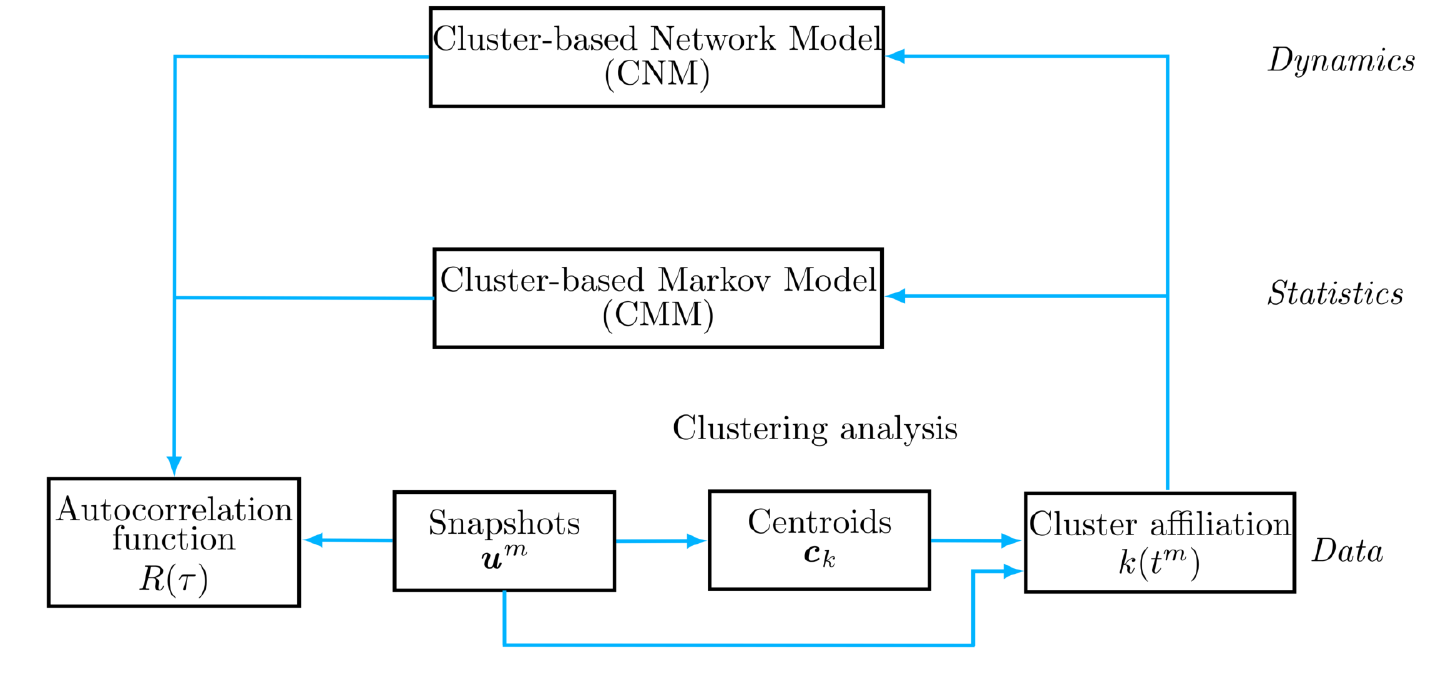}
	\caption{Principle sketch of cluster-based modelling. 
                The time-resolved snapshots are 
		partitioned into a predetermined number of clusters represented by centroids in an unsupervised manner.
                Thereafter, each snapshot has a cluster-affiliation $k(t^m)$ being the index of the closest centroid.
                Cluster-based Markov models (CMM) describe the evolution of the population of these clusters.
                The solution of CMM quickly converges against the asymptotic probability distribution.
                The proposed cluster-based network model (CNM)  
                resolves the dynamic transitions between the clusters by a deterministic-stochastic network.}
	\label{Fig:Methodology}
\end{figure}

In this section, 
we propose a novel cluster-based reduced-order model (ROM)
for the coherent structure dynamics starting at the time-resolved snapshots. 
In \S~\ref{ToC:Method:Clustering} and \S~\ref{ToC:Method:MarkovModel}
clustering and cluster-based Markov models (CMM) are recapitulated.
Section \ref{ToC:Method:NetworkModel} proposes a novel data-driven dynamic network
resolving the transition dynamics between the clusters.
In \S~\ref{ToC:Method:Computation}, 
the time-discrete cluster-based ROM 
is enhanced for a continuous-time velocity prediction.  
The model validation includes the autocorrelation function of the flow 
as discussed in \S~\ref{ToC:Method:Validation}.
Figure \ref{Fig:Methodology} previews the methodology
and will be explained later in the section.
The relative advantages of CMM and CNM 
are illustrated for the Lorenz attractor
in \S~\ref{ToC:Method:Lorenz}.

\begin{figure}
	\centering
	\def\svgwidth{0.7\linewidth}
	 \includegraphics[width=0.8\linewidth]{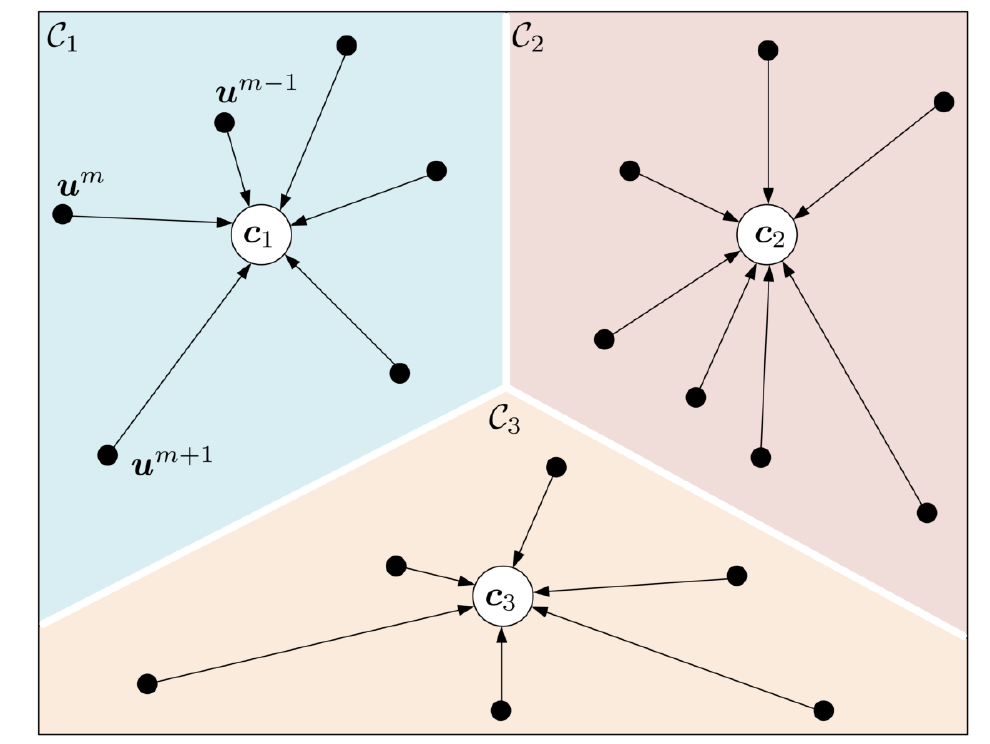}
	\centering
	\caption{
	Clustering exemplified in a two-dimensional state space. 
	The snapshots $\left \{ \bm{u}^m \right \}_{m=1}^M$ are coarse-grained into $3$  clusters. 
	Each centroid $\bm{c}_{k}, k=1, 2 ,3$, 
	is the averaged field of all snapshots belonging to this cluster.
	Every centroid $\bm{c}_k$  is associated with a Voronoi cell $\mathcal{C}_k$, 
	i.e., a region in which the points are closer to $\bm{c}_k$ than any other centroid.
	The cluster affiliation for a snapshot is the cluster index of the closest centroid which has been indicated by an arrow.
	By definition, the cluster affiliation is the index of the corresponding Voronoi cell.}
	\label{fig:ClusteringSketch}
\end{figure}

\subsection{Clustering as coarse-graining}
\label{ToC:Method:Clustering}
We consider velocity fields in a steady domain $\Omega$
which may be  obtained from experiments or from  numerical simulations. 
Starting point is an ensemble of $M$ statistically representative, time-resolved snapshots
as employed for cluster-based models \citep{Kaiser2014jfm},
Dynamic Mode Decomposition (DMD)  \citep{Rowley2009jfm,Schmid2010jfm} 
or Proper Orthogonal Decomposition (POD) \citep[see, e.g.,][]{Holmes2012book}.
The velocity field is equidistantly sampled  with time step $\Delta t$,
i.e., the $m$th instant reads $t^{m}=m \Delta t$.
The corresponding snapshot velocity field is denoted by
$\bm{u}^{m}\left(\bm{x} \right) :=\bm{u}\left(\bm{x}, t^{m}\right)$, 
$m=1,\ldots,M$.

Cluster analysis lumps  similar objects into clusters.
This lumping of data   is performed  in an unsupervised manner,
i.e., no advance labeling or grouping of the data has been performed.
In cluster-based models, 
the $M$ snapshots $\bm{u}^{m} ( \bm{x} )$ are 
coarse-grained into $K$ \emph{clusters}
represented by the \emph{centroids} $\bm{c}_k (\bm{x})$, $k=1,\ldots,K$, 
using the unsupervised k-means++ algorithm
\citep{Steinhaus1956,MacQueen1967proc,Lloyd1982ieee}. 
The centroids characterise the typical flow patterns of each clusters, also called modes
in the ROM community.
The corresponding \emph{cluster-affiliation function} maps 
a velocity field $\bm{u}$
to the index of the closest centroid,
\begin{equation}
k (\bm{u})  = \arg\min_i \Vert \bm{u} - \bm{c}_i \Vert_{\Omega},
\label{Eqn:ClusterAffiliation}
\end{equation}
where $ \Vert \cdot \Vert_{\Omega} $ denotes the standard Hilbert space norm 
in the domain $\Omega$ (see appendix \ref{ToC:Method:POD}).
This function defines cluster regions as Voronoi cells around the centroids
\begin{equation}
{\cal C}_i = \left \{  \bm{u} \in {\cal L}^2 ( \Omega )   \colon  k (\bm{u}) = i \right \}.
\label{Eqn:ClusterRegion}
\end{equation}
This function can also be employed to map the snapshot index $m$
to the representative cluster index 
$ k(m) := k\left ( \bm{u}^m \right ) $.
Alternatively, the characteristic function
\begin{equation}
\chi_i^m :=\left\{\begin{array}{ll}{1,} & \text{if} \quad i=k(m) \\ 
                                       {0,} & \text{otherwise}
                      \end{array}\right.
\label{Eqn:CharacteristicFunction}
\end{equation}
describes if the $m$th snapshot is affiliated with the $l$th centroid.
The latter two quantities are equivalent.

The performance of a set of centroids 
$\left \{ \bm{c}_k \right \}_{k=1}^K$
with respect to a given set of snapshots  
$\left \{ \bm{u}^m \right \}_{m=1}^M$
is measured
by the average variance of the snapshots with respect to their closest centroid
The corresponding \emph{inner-cluster variance} reads
\begin{equation}
J \left( \bm{c}_1,\ldots,\bm{c}_K \right ) 
= \frac{1}{M} \sum_{m=1}^{M} \left \Vert \bm{u}^{m}-\bm{c}_{k(m)} \right \Vert^2_\Omega .
\label{Eqn:ClusterVariance}
\end{equation}
The optimal centroids
$\left \{ \bm{c}_k^\star \right \}_{k=1}^K$ 
minimize this inner-cluster variance,
\begin{equation}
\left ( \bm{c}_1^\star, \ldots, 
         \bm{c}_K^\star 
\right )
=\underset{\bm{c}_1,\ldots,\bm{c}_K}{\arg\min}  \> J \left( \bm{c}_1,\ldots,\bm{c}_K \right ) 
\label{Eqn:KMeans}
\end{equation}
The argument is indeterminate with respect to a re-ordering.
For CMM, we chose the first cluster as the one with the highest population, 
i.e.\ the largest number of associated snapshots.
The $(k+1)$th cluster, $k>1$, 
has the largest transition probability from the $k$th one.

The optimization problem  \eqref{Eqn:KMeans}
is solved by the k-means++ algorithm.
The $K$ centroids are initialized randomly and then iterated 
until convergence is reached or when the variance $J$ is small enough. 
k-means++ repeats the clustering process 30 times and take the best set of centroids.

The number of snapshots $n_k$ in cluster $k$ is given by
\begin{equation}
n_k=\sum_{m=1}^{M} \chi_k^{m}.
\end{equation}
The centroids are the mean velocity field of all snapshots in the corresponding cluster.
In other words, 
\begin{equation}
\bm{c}_k = \frac{1}{n_k} \sum_{\bm{u}^{m} \in \mathcal{C}_k} \bm{u}^{m}=\frac{1}{n_k} \sum_{m=1}^{M} \chi_k^{m} \bm{u}^{m}.
\label{Eqn:centroids}
\end{equation}
In the following centroid visualizations, 
we accentuate  the vortical structures 
by displaying the fluctuations $\bm{c}_k - \overline{\bm{u}}$
around the snapshot mean $\overline{\bm{u}}$
and not the full velocity field $\bm{c}_k$.
\subsection{Cluster-based Markov model (CMM)}
\label{ToC:Method:MarkovModel}

We briefly recapitulate CMM by \citet{Kaiser2014jfm} as our benchmark cluster-based reduced-order model.
In CMM, the state variable is the cluster population 
$\bm{p}=\left[ p_1,\ldots,p_K \right ]^{\mathrm{T}}$, 
where $p_i$ represents the probability to be in cluster $i$
and the superscript $\mathrm{T}$ denotes the transpose.
The transition between clusters in a given time step $\Delta t^c$
is described by the transition matrix ${\sf{\bm{P}}} = (P_{ij}) \in {\cal R}^{K \times K}$.
The superscript `$c$' refers to cluster-based model.
Here, $P_{ij}$ is the transition probability to move from cluster $j$ to cluster $i$.
Let $\bm{p}^l$ be the probability vector at time $t^l = l \Delta t^c$,
then the change in one time step is described by 
\begin{equation}
\bm{p}^{l+1} = {{\sf{\bm{P}}}} \> \bm{p}^l
\label{Eqn:CMM}
\end{equation}
With increasing iterations, the iteration \eqref{Eqn:CMM}
converges to the asymptotic probability 
$\bm{p}^{\infty} := \lim\limits_{l\to \infty} \bm{p}^l$.
In a typical case, \eqref{Eqn:CMM} has a single fixed point $\bm{p}^{\infty}$.
For completeness, a continuous form of Markov models 
with new transition matrix ${{\sf{\bm{P}}}}^c$ is mentioned: 
\begin{equation}
\frac{d\bm{p}}{dt}={\sf{\bm{P}}}^{c} \> \bm{p}.
\label{Eqn:CMMt}
\end{equation}
From the time-continuous form \eqref{Eqn:CMMt}, 
the time-discrete one \eqref{Eqn:CMM} can be derived.
The opposite is not generally true.
In the following,
no continuous Markov models are employed.

A CMM of the time-resolved snapshots
starts with cluster affiliation \eqref{Eqn:ClusterAffiliation}
which  can also be considered  function of time $k(t)$.
We refer to the original paper for the determination of $P_{ij}$ from $k(t)$.
The time step $\Delta t^c$ is a critical design parameter for CMM.
A good choice is a value where the transition 
from one cluster to the next is likely.
If the time step is too small, 
the Markov model idles many times in each cluster 
for a stochastic number of times before transitioning to the next cluster.
The model-based transition time may thus significantly deviate 
from the deterministic data-driven trajectories through the clusters.
If the time step is too large, one may miss intermediate clusters.
We choose $\Delta t^c= T/10$, where $T$ is the characteristic period of the flow.
On a circular limit cycle with uniform rotation,
this value is optimal  for $K=10$ clusters,
enforcing the transition from one cluster 
to the next in each time step.

\begin{figure}
	\centering
	\sidecaption{subfig:a}
	\raisebox{-\height}{\def\svgwidth{0.2\linewidth}\includegraphics[width=0.45\linewidth]{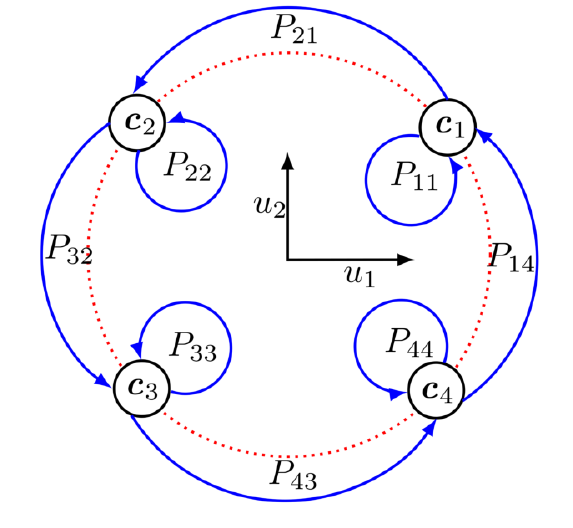}}%
	\hfill
	\sidecaption{subfig:b}
	\raisebox{-\height}{\def\svgwidth{0.2\linewidth}\includegraphics[width=0.45\linewidth]{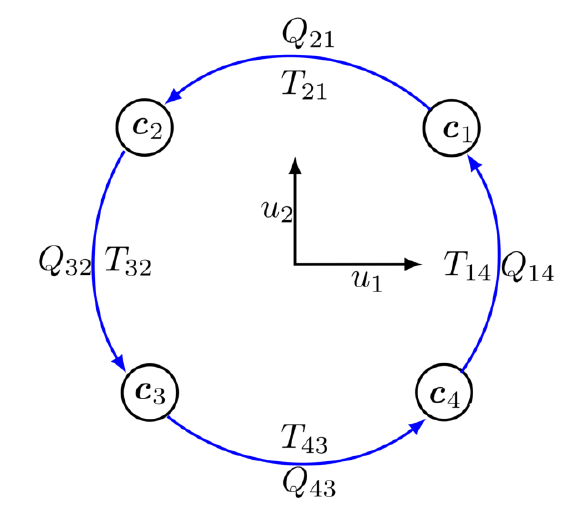}}
	\par
	\sidecaption{subfig:c}
	\raisebox{-\height}{\def\svgwidth{0.2\linewidth}\includegraphics[width=0.45\linewidth]{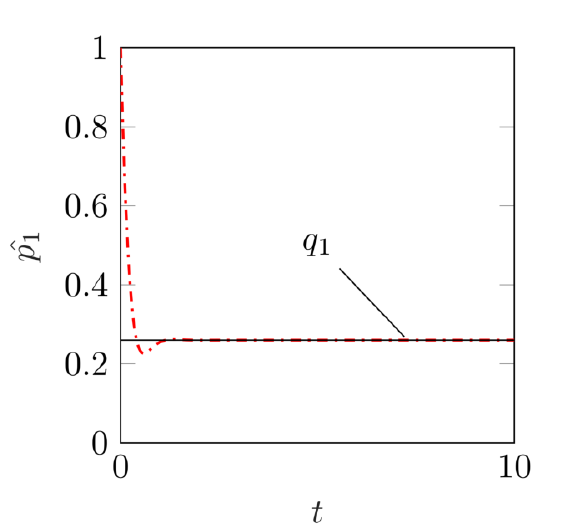}}%
	\hfill
	\sidecaption{subfig:d}
	\raisebox{-\height}{\def\svgwidth{0.2\linewidth}\includegraphics[width=0.45\linewidth]{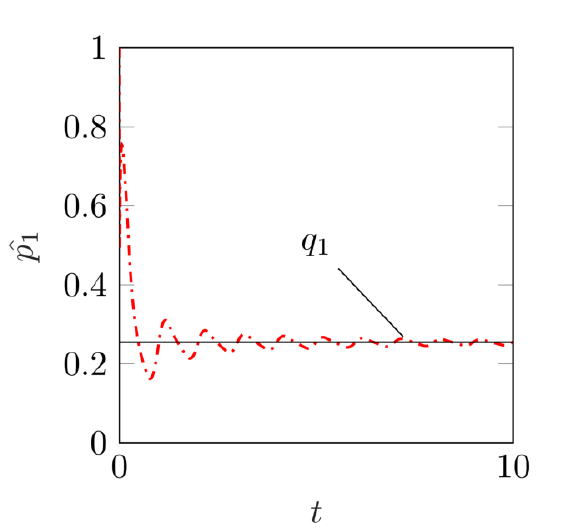}}%
	\par
	\sidecaption{subfig:e}
	\raisebox{-\height}{\def\svgwidth{0.2\linewidth}\includegraphics[width=0.45\linewidth]{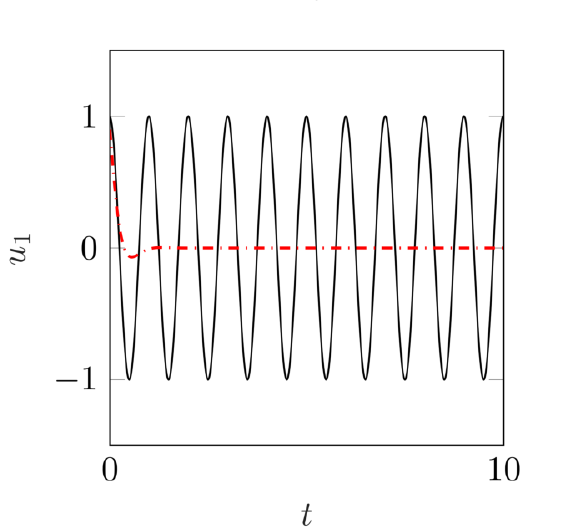}}%
	\hfill
	\sidecaption{subfig:f}
	\raisebox{-\height}{\def\svgwidth{0.2\linewidth}\includegraphics[width=0.45\linewidth]{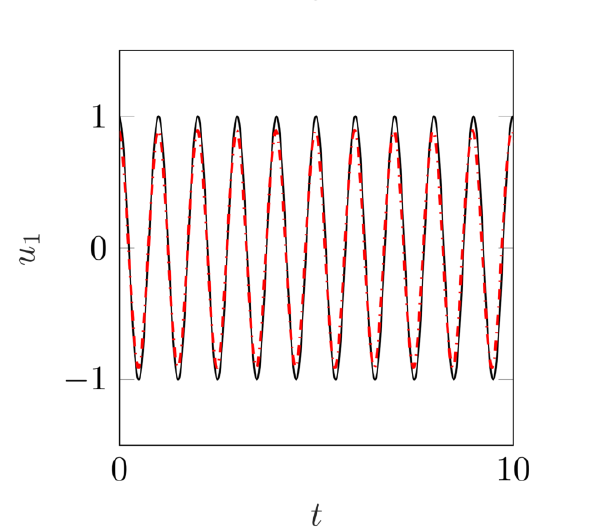}}%
	\caption{Introduction of cluster-based models for a limit cycle example. 
                 CMM and CNM are displayed in the left and right column, respectively.
                 The uniform rotation $u_{1}=\cos(2\pi t)$, $u_{2}=\sin(2\pi t)$ 
                 in a two-dimensional plane is discretized by $4$ centroids $\bm{c}_{k}$, $k=1,...4$.
        \subref{subfig:a} Phase portrait of CMM with time step $\Delta t=1/16$. 
            The centroids are near the limit cycle (red dashed line).	 
	    The state vector residing in centroid $\bm{c}_{i}$ 
            has the probability $P_{ii}$ to stay in its state 
            and $P_{ji}$ to transition to centroid $\bm{c}_{j}$ in the considered time-step.
        \subref{subfig:b} Phase portrait of the CNM. The state in centroid $\bm{c}_{i}$ 
            moves uniformly to its counter-clockwise neighbour taking a quarter period $T_{14}=T_{21}=T_{32}=T_{43}=1/4$.
            Here, $Q_{14}=Q_{21}=Q_{32}=Q_{43}=1$ and $Q_{ij}=0$ otherwise.
            The estimated probability evolution starting in cluster $i=1$ at $t=0$ is illustrated for CMM \subref{subfig:c} and CNM \subref{subfig:d}. 
            (\subref{subfig:e}) and \subref{subfig:f} present the model-based evolution of the first coordinate $u_{1}$ for CMM and CNM, respectively.
            In \subref{subfig:c}--\subref{subfig:f}, the solid black lines correspond to the uniform rotation 
            and the dashed red line to the model.}
	\label{Fig:IllustrationCase}
\end{figure}
In figure \ref{Fig:IllustrationCase} (left column),
the effect of the suboptimal time step is illustrated 
for the CMM of a  uniform rotation
$ u_1 = \cos ( 2\pi t)$, $u_2 = \sin (2 \pi t)$.
Here, 4 clusters and a time step $\Delta t^c = 1/16$ are chosen.
The probability to stay in the cluster during one time step is $P_{11}=P_{22}=P_{33}=P_{44}= 3/4$ and 
the transition probability to the next counter-clockwise neighbour is  $P_{14}=P_{21}=P_{32}=P_{43}= 1/4$.
Thus, the probability to stay in one cluster for $l$ steps exponentially decays, $P_{11}^l$.
In contrast,  the uniform rotation commands 
that the state is exactly three time steps 
in one cluster before it leaves in the fourth step.
This example motivates the proposed cluster-based reduced-order model,
foreshadowed in figure \ref{Fig:IllustrationCase} (right column)
and explained in the following section.

\subsection{Cluster-based network model (CNM)}
\label{ToC:Method:NetworkModel}

For CMM, the time step $\Delta t^c$ is, as mentioned, an important design parameter.
This design parameter can be avoided by the new proposed \emph{Cluster-based Network Model} (CNM).
The key idea is to abandon the `stroboscopic' view of CMM 
and focus on non-trivial transitions from cluster $j$ to cluster $i$.
These transitions are characterized by two parameters:
the probability $Q_{ij}$ and a time-scale $T_{ij}$.
Evidently, no time-step is needed for the description 
and the assumption of a constant transition time is found 
to be much more aligned with shear flow modes than assuming an exponential decay of residence time.
Moreover, it could be relaxed by assuming a probability distribution of transition times.

In the following, the transition probability and transition time
are inferred from the cluster affiliation function $k(t)$.
The continuous form is convenient for discussion.
The time discrete affiliation function $k(m)$ can be made continuous 
by taking the cluster of the snapshot which is closest in time,
$$ k(t) = k \left ( \arg\min_m 
            \left \vert t - t_m 
            \right \vert 
             \right ). $$

\begin{figure}
	\centering
	\def\svgwidth{0.8\linewidth}
	\includegraphics[width=0.6\linewidth]{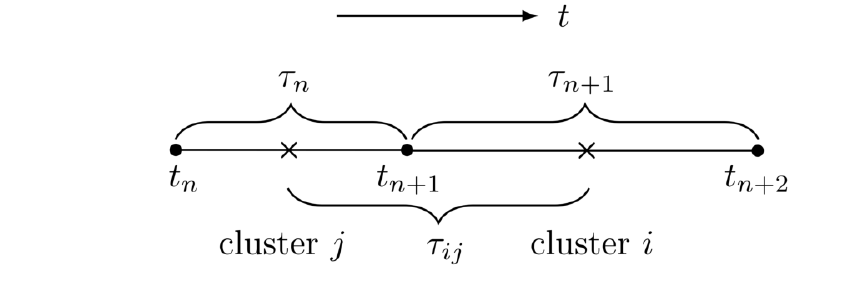}
	\caption{Sketch of times and periods employed in the cluster-based network model. 
                 $\times$ marks the center of the cluster residence time, 
while $\bullet$ denotes the transition between clusters.}
	\label{fig:TimeSketch}
\end{figure}

The $n$th transition time $t_n$ of the cluster affiliation 
is recursively defined as the first discontinuity of $k(t)$ for $t>t_{n-1}$.
Here, $t_0 = 0$.
The transition time $t_n$ satisfies 
\begin{equation}
k \left ( t_{n}-\varepsilon \right ) \neq 
k \left ( t_{n}+\varepsilon \right )
\label{Eqn:TransitionCriterion}
\end{equation}
for any sufficiently small positive $\varepsilon$.
For $t \in \left( t_n, t_{n+1} \right )$,
the data-based trajectory is assumed to stay in cluster $k$
at the averaged time $(t_{n+1}+t_n)/2$ (see figure \ref{fig:TimeSketch}).
The residence time in this cluster is defined by 
 \begin{equation}
 \tau_{n} :=t_{n+1}-t_{n}.
\label{Eqn:ResidenceTime1}
 \end{equation}
Let $j$ and $i$ be the indices of the clusters
after $t_n$ and $t_{n+1}$ respectively.
Then the transition time from $j$ to $i$ 
is defined as half of the residence time of both clusters,
\begin{equation}
\tau_{ij} := \frac{\tau_{n}+\tau_{n+1}}{2} = \frac{t_{n+2}-t_n}{2}.
\label{Eqn:ResidenceTime2}
\end{equation}
This definition may appear arbitrary but is the least-biased guess consistent with the available data.
The sum of all residence times from a given data set add up to the total investigated time period $T_0$.

The direct transition probability $Q_{ij}$ and transition time $T_{ij}$
can be inferred from the data.
Then,
\begin{equation}
Q_{ij}=\frac{n_{ij}} {n_{j}}, i,j=1,\ldots,K;
\label{Eqn:DTM}
\end{equation}
where $n_{ij}$ is the number of transitions  from $\bm{c}_{j}$ to $\bm{c}_{i}$
and $n_j$ the number of transitions departing from $\bm{c}_{j}$ regardless of the destination point,
\begin{equation}
n_j=\sum_{i=1}^K n_{i j}, i,j=1,\ldots,K.
\end{equation} 
We emphasize that $n_{ii}=0$ for $i=1,\ldots,K$ 
by very definition of a direct transition.
The \emph{direct transition matrix} (DTM) ${\sf{\bm{Q}}} = \left (Q_{ij} \right ) \in {\cal R}^{K \times K}$
lumps these probabilities into a single entity.

Similarly, the direct transition time $T_{ij}$
from cluster $j$ to cluster $i$
is taken to be the average of all values.
This average is symbolically denoted by
\begin{equation}
T_{i j}= \langle  \tau_{ij}  \rangle
\label{Eqn:TransitionTime}
\end{equation}
These values are lumped into the matrix ${\sf{\bm{T}}} = \left (T_{ij} \right ) \in {\cal R}^{K \times K}$.

It should be noted that a given trajectory 
may repeatedly pass through the same clusters (Voronoi cells) 
with different residence and transition times.
With enough data this variability may be incorporated into the model.
Our goal is to compare the Markov model with the most simple network model
where constant (averaged) transition times are assumed.

CNM predicts the asymptotic cluster probability $p^\infty_i$ in cluster $i$.
Let $[0,T_0]$ be a sufficiently long time horizon simulated by the model.
Then, the probability to stay in cluster $i$ is the cumulative residence time
normalized by the simulation time,
\begin{equation}
p^\infty_i= \frac{\sum \tau_i}{T_0}.
\label{Eqn:CNM:pi}
\end{equation}

We return to the introductory example depicted in figure \ref{Fig:IllustrationCase}.
The CNM is seen to accurately describe the uniform rotation (subfigure f)
and  correctly yields the asymptotic cluster probability $p_i^\infty =1/4$, $i=1,\ldots,4$.
In contrast, the prediction horizon of the CMM is limited to roughly one period.
After this time, the initial condition is forgotten and the asymptotic distribution is 
reached---rendering CMM unsuitable for dynamic prediction.
However, CMM predicts the asymptotic state faster than the CNM.
For this particular example, 
the CMM could be made equivalent to the CNM by choosing $\Delta t^c=1/4$.
However, the Markov model will inevitably diffuse 
the state with a range of cluster-transition times,
e.g.\ for non-uniform rotation or more complex dynamics.

\subsection{Velocity fields associated with the cluster-based reduced-order models}
\label{ToC:Method:Computation}

The CMM describes the cluster popoulation 
\begin{equation}
\label{Eqn:ProbabilityVector}
\bm p=\left[ p_{1},...,p_{K} \right]^T
\end{equation}
at discrete times $t=l \> \Delta t^c$.
In the following, 
this population is considered to be continuous in time,
e.g.\ by using  linear or higher-order interpolation.
The corresponding velocity field  $\bm u(\bm x ,t)$ at time $t$ 
is defined as the expectation value,
\begin{equation}
\bm u(\bm x ,t) =\sum_{i=1}^{K} p_{i}(t) \> \bm c_{i}(\bm x)
\label{Eqn:EvolutionVelocity}
\end{equation}
where $\bm c_{i}$ is the $i$th centroid.

The CNM is based on centroid visits at  discrete times.
The clusters $k_0$, $k_1$, $k_2$, $\ldots$ 
are visited at times 
\begin{equation}
 t_0=0,  \quad  
   t_1=T_{k_1 k_0}, \quad 
   t_2= t_1 + T_{k_2 k_1}, \ldots 
\label{Eqn:CNM:Times}
\end{equation}
consistent with the direct transition matrix $(Q_{ij})$ 
and the transition times $T_{ij}$.
A uniform motion is assumed between these visits.
In other words,  
for  $t \in [t_n,t_{n+1}] $ the velocity field reads 
\begin{equation}
\bm{u}(\bm{x},t) = \alpha_{n} (t) \> \bm{c}_{k_n} (\bm{x})  
              + \left [ 1-\alpha_n(t) \right ] \> \bm{c}_{k_{n+1}} (\bm{x}),
\quad \alpha_n = \frac{t_{n+1}-t}{t_{n+1}-t_n} .
\label{Eqn:CNM:Velocity}
\end{equation}
We note that a smoother motion may be achieved with splines.

The actual flow computations are based 
on a lossless proper orthogonal decomposition (POD),
as elaborated in the appendix \ref{ToC:Method:POD}.
The interpolations are performed with the mode amplitudes 
$\bm{a}=\left [ a_1,\ldots,a_N \right]^T$
before transcribed into velocity fields via the POD expansion.

\begin{figure}[htb]
	\includegraphics[width=\linewidth]{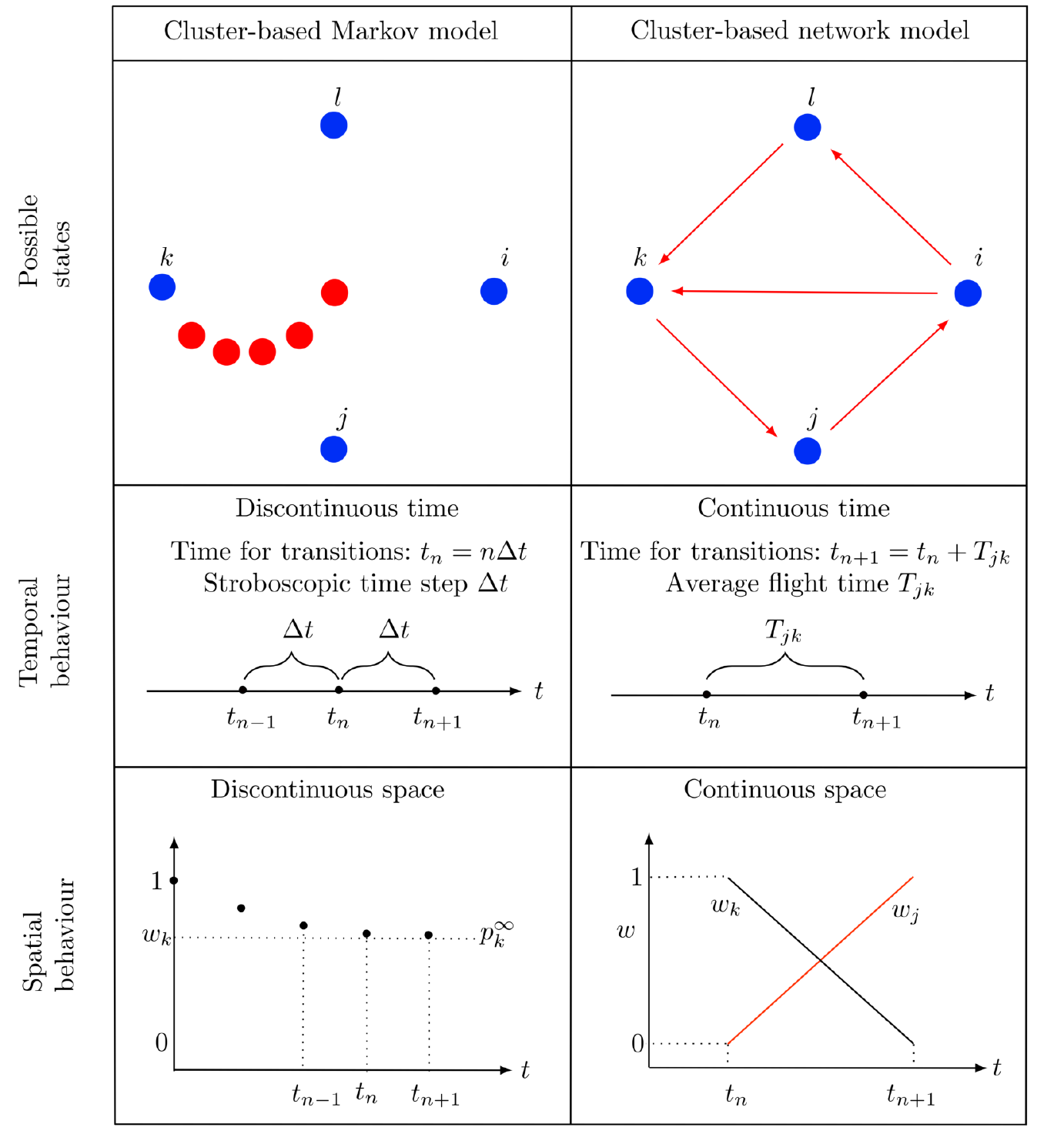}
	\caption{Comparison of the cluster-based Markov  (CMM) and  network model (CNM).
	        For reasons of simplicity, an example with four centroids $i$, $j$, $k$ and $l$ is shown.
	        The CNM exemplifies the evolution of the weight of centroid $k$.
	        For the CMM, the transition from cluster $k$ to $j$ is shown.
	        The predicted state  \eqref{Eqn:CNM:Velocity} is determined by 
	        the weights $w_k(t) = \alpha(t)$ and $w_j = 1-\alpha(t)$, $\alpha=(t_{n+1}-t)/T_{jk}$ 
	        in the time interval $\left[ t_n,t_{n+1} \right]$. For details see text.}
	\label{fig:Comparison_CNM_CMM}
\end{figure}
Figure \ref{fig:Comparison_CNM_CMM} 
compares the classical CMM  with stroboscopic temporal prediction of discrete states 
and the proposed CNM with time-continuous uniform motion 
on a network of routes between two centroids.
In the top row, the possible states are illustrated.
In case of the CMM, the states \eqref{Eqn:EvolutionVelocity},
denoted by red dots,  
quickly converge to the mean flow, like RANS simulations.
The CNM-predicted state  \eqref{Eqn:CNM:Velocity}
moves on the directed network, marked by red arrows, 
and is reminiscent of large-eddy simulations.
As displayed in the middle row, 
the CMM is discrete in time 
while the CNM dynamics is time continuous 
with cluster visits after pre-specified transition times $T_{jk}$.
The bottom row shows another difference: 
CMM describes averages over all centroids 
while the CNM only allows for linear interpolations 
between two neighbouring centroids.
This interpolation is  consistent 
with the purpose to accurately resolve evolving coherent structures.
Averaging over many centroids acts like a low-pass filter
mitigating the fluctuation level and thus underresolving the coherent structures.
 
\subsection{Validation of the cluster-based reduced-order models}
\label{ToC:Method:Validation} 
Following \citet{Protas2015jfm},  
the cluster-based model is validated based on the computed and predicted
autocorrelation function of the velocity field.
The unbiased autocorrelation function reads 
\begin{equation}
R(\tau) :=\frac{1}{T-\tau} \int_{\tau}^T \int_{\Omega}  \bm{u}(\bm{x}, t-\tau) \cdot \bm{u}(\bm{x}, t)  \>  d\bm{x} \> dt, \quad \tau \in[0, T).
\end{equation}
This function reveals the turbulent fluctuation level $R(0)$ 
and the frequency spectrum.
Moreover, the problem of comparing two trajectories 
with finite dynamic prediction horizons 
due to the increasing phase mismatch is avoided \citep{Pastoor2005cdc}.

In case of the CNM, the modeled autocorrelation function $\hat{R}$
is based on the modelled velocity field \eqref{Eqn:CNM:Velocity}.
In case of the CMM, 
the time integration quickly leads to the average flow 
and is not indicative for the range of possible initial conditions.
Hence, $K$ trajectories are considered starting with $p_k=1$ for each cluster $k$, 
or, equivalently, $\bm{p}^{\circ k} (t=0)= \left[ \delta_{1k}, \ldots, \delta_{Kk} \right]^{\mathrm{T}}$.
These cluster-specific autocorrelation functions are weighted with the cluster probability $p_i^\infty$
\begin{equation}
\hat{R} (\tau) :=\sum\limits_{k=1}^K  p_i^\infty \> 
\int_{0}^T \int_{\Omega}  \bm{u}^{\circ k} (\bm{x}, t) \cdot \bm{u}^{\circ k} (\bm{x}, t+\tau ) \> d\bm{x} \> d t, \quad \tau \in[0, T),
\end{equation}
where $\bm{u}^{\circ k}$ denotes the CMM-predicted velocity field starting in cluster $k$.

\subsection{Lorenz system as an illustrating example}
\label{ToC:Method:Lorenz}

Following the original CMM paper by \citet{Kaiser2014jfm},
the CNM is illustrated for the celebrated \citet{Lorenz1963jas} system,
arguably the first demonstration of chaotic dynamics in low-dimensional dynamics.
The Lorenz system is a three-dimensional autonomous system 
of nonlinear ordinary differential equations.
The derivation was inspired by a Galerkin model of Rayleigh-Benard convection,
but typically-selected parameters clearly 
exceed the range of model validity \citep{Sparrow1982book}.
The system features non-periodic, deterministic, dissipative dynamics
associated with exponential divergence and convergence to a fractal strange attractor.
The three coupled nonlinear differential equations read 
\begin{subequations}
\label{Eqn:LorenzSystem}
\begin{eqnarray} 
\frac{\mathrm{d} x}{\mathrm{d} t} &=&\sigma(y-x) \\ 
\frac{\mathrm{d} y}{\mathrm{d} t} &=&x(r-z)-y \\ 
\frac{\mathrm{d} z}{\mathrm{d} t} &=&x y-b z 
\end{eqnarray}
\end{subequations}
with the system parameters $\sigma=10$, $b=8/3$ and $r=28$. 
For these parameters,
there are three unstable fixed points at $(0,0,0)$ 
and $(\pm \sqrt{72},\pm \sqrt{72},27)$, denoted by $F^+$ and $F^-$, respectively. 
The attractor of Lorenz system resembles two butterfly wings around $F^+$ and $F^-$in phase space. 
The trajectory typically oscillates for several periods with increasing amplitude 
around a fixed point ($F^+$ or $F^-$)
before it moves to the other wing. 
The number of revolutions made on either 
side varies unpredictably from one cycle to the next.

\begin{figure}[htb]
	\centering
	\sidecaption{subfig:a}
	\raisebox{-\height}{\def\svgwidth{0.2\linewidth}\includegraphics[width=0.44\linewidth]{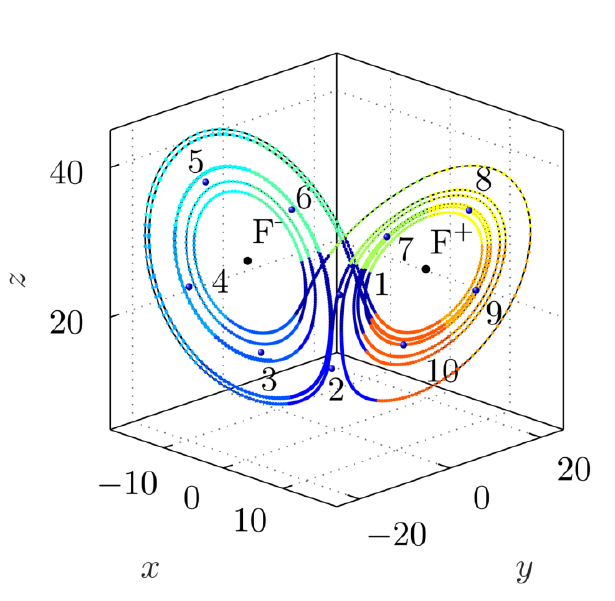}}%
	\hfill
	\sidecaption{subfig:b}
	\raisebox{-\height}{\def\svgwidth{0.2\linewidth}\includegraphics[width=0.46\linewidth]{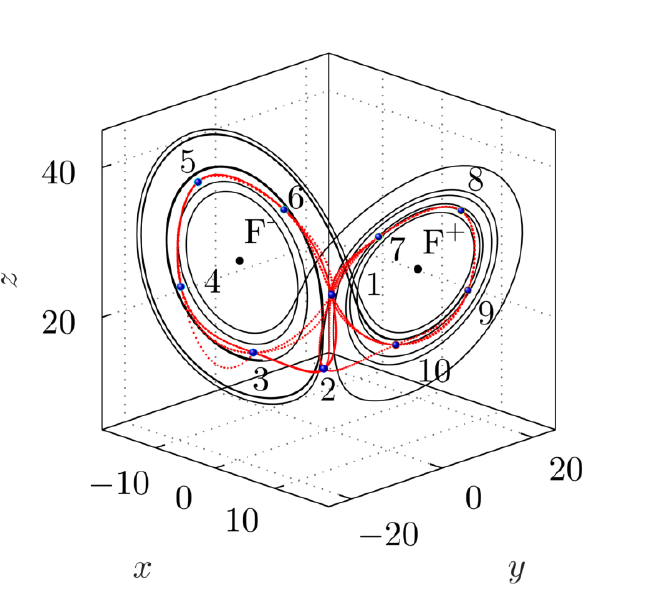}}%
	\caption{Cluster-based network model (CNM) for the Lorenz attractor.
                \subref{subfig:a} Cluster partitioning: 
                The centroids are displayed as colored solid circles. 
                A trajectory is illustrated by a black curve.
                The dots on this trajectory are colored according to their cluster affiliation. 
                The clusters $k=3,4,5,6$ oscillate around the fixed point $F^-$
		and clusters $k=7,8,9,10$ around $F^+$. 
                The clusters $k=1,2$ connect both `ears' of the Lorenz attractor. 
                \subref{subfig:b} The trajectory of the CNM (red dashed line). 
		The centroids represent the network nodes and edges represent possible transitions.
		Here,  trajectory of CNM is obtained by a spline-interpolation through the visited centroids.}
		\label{fig:LorenzTrajectory}
\end{figure}

\begin{figure}
	\centering
	\sidecaption{subfig:a}
	\raisebox{-\height}{\def\svgwidth{0.445\linewidth}
	\includegraphics[width=0.44\linewidth]{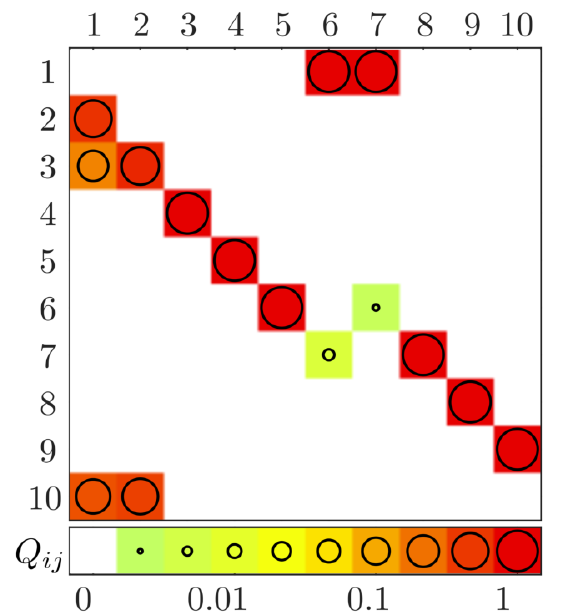}}
	\hfill
	\sidecaption{subfig:b}
	\raisebox{-\height}{\def\svgwidth{0.445\linewidth}
	\includegraphics[width=0.44\linewidth]{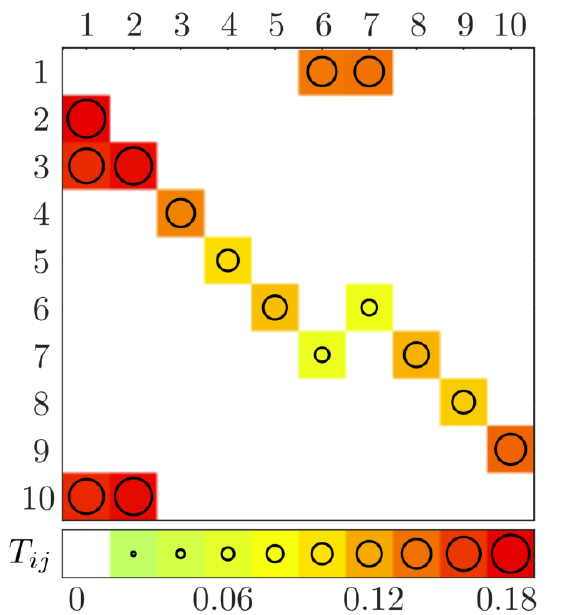}}
	\caption{Cluster-based network model of the Lorenz attractor.
	\subref{subfig:a} Direct transition matrix $(Q_{ij})$ 
        \subref{subfig:b} Transition time $(T_{ij})$. 
	\label{fig:DirectTransitionMatrixLorenz}}
\end{figure}

The Lorenz equations (\ref{Eqn:LorenzSystem}) are solved 
employing an explicit fourth-order Runge-Kutta scheme 
with an initial condition on the attractor.
The time-resolved snapshots data  $\bm{x}(t_{m})$ with $\bm x=[x,y,z]^T$ 
are collected at a sampling time step $\Delta t=0.005$
corresponding roughly to one thousands of a typical oscillation period.
The k-means ++ algorithm partitions $M=1,000,000$ snapshots into $K=10$ clusters. 
Figure \ref{fig:LorenzTrajectory}\subref{subfig:a} 
displays a phase portrait of the corresponding clusters.
The snapshots associated with one cluster are highlighted by the same color. 
The $10$ clusters feature three different subsets:
the transition clusters $k=1,2$ between two butterfly wings,
the $F^-$ wing related cluster $k=3,4,5,6$ 
and clusters $k=7,8,9,10$ associated with the $F^+$ wing. 
The wing-related cluster groups 
represent approximately $90^\circ$ phase bins
and don't resolve the amplitude. 
Evidently, the 10 clusters are coarse representations of the state.

In the following, the dynamics are resolved 
by the network model of \S~\ref{ToC:Method:NetworkModel}. 
The $10$ centroids are considered as nodes in the network. 
The transition between these centroids define directed edges 
characterised by direct transition matrix  ${\sf{\bm{Q}}}$
and the flight times ${\sf{\bm{T}}}$.
The connectivity is described by the adjacency matrix
$H({\sf{\bm{Q}}})$ where $H$ denotes the Heaviside function: non-vanishing 
elements of ${\sf{\bm{Q}}}$ are replaced by unity  \citep{newman2010networks}. 
Figure \ref{fig:DirectTransitionMatrixLorenz}, 
displays the DTM ${\sf{\bm{Q}}}$ (subfigure (a)) 
and associated transition time matrix ${\sf{\bm{T}}}$ (subfigure (b)).
The matrices
reveal three distinct cluster groups
consistent with the phase diagram of figure \ref{fig:LorenzTrajectory}.
Clusters $1$ and $2$ allow transitions to $3$ and $10$,
i.e., the $F^-$ and $F^+$ wing, respectively,
and have  been called \emph{flipper clusters} by \citet{Kaiser2014jfm}.
The cluster transition sequence
 $3 \to 4 \to 5 \to 6 \to 1 \to 2 \to 3$ 
is the dominant cyclic group associated with the $F^-$ wing.
Another cyclic groups skips cluster $2$:  
$3 \to 4 \to 5 \to 6 \to 1 \to 3$.
A cyclic group through the $F^+$ wing reads
$10 \to 9 \to 8 \to 7 \to 1 \to 10$.
A longer sequence includes the $2$nd cluster: 
$10 \to 9 \to 8 \to 7 \to 1 \to 2 \to 10$.
The transition times in the wing centroids are noticeably smaller
than the passage through the flipper clusters.

\begin{figure}
	\centering
	\def\svgwidth{0.8\linewidth}
	\includegraphics[width=0.7\linewidth]{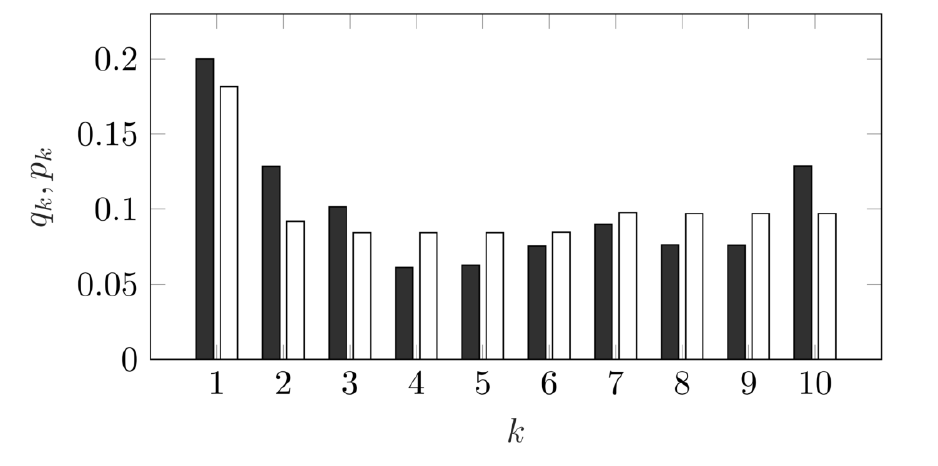}
	\caption{Cluster probability distribution 
                 of the Lorenz system (solid rectangles) 
                 and the corresponding cluster-based network model (open rectangles).
                 The model results are based on $20,000$ transitions.}
	\label{fig:ProbLorenzCNM_Data}
\end{figure}

\begin{figure}[htb]
	\centering
	\def\svgwidth{0.8\linewidth}
	\includegraphics[width=0.8\linewidth]{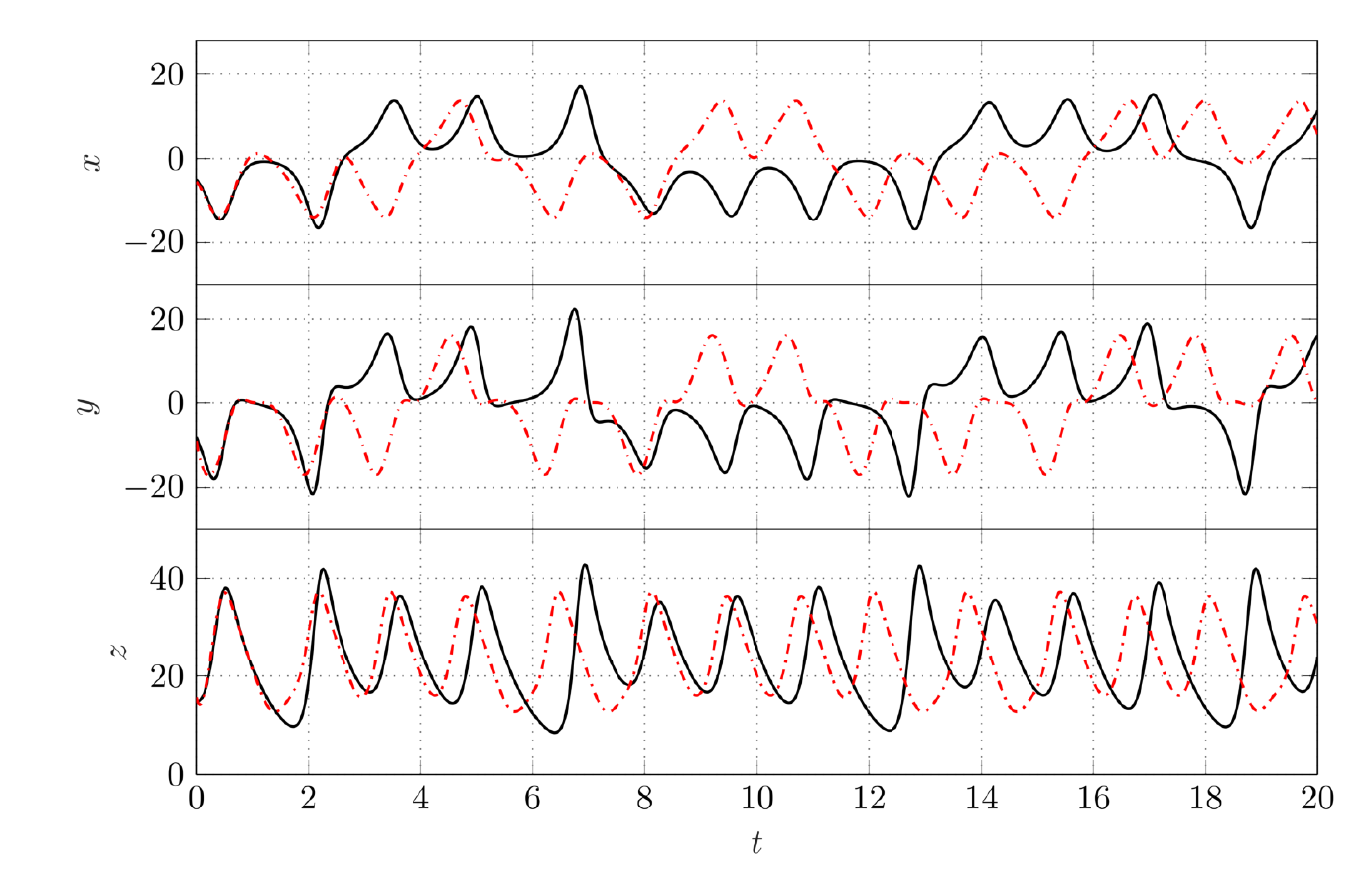}
	\caption{Evolution of the Lorenz system $x$, $y$, $z$, $t\in[0,20]$
                 from integrating the dynamical system (black solid line) 
                 and from the prediction of the cluster-based network model (red dashed line).}
	\label{fig:ModeEvolution}
\end{figure}

\begin{figure}[htb]
	\centering
	\def\svgwidth{0.8\linewidth}
	\includegraphics[width=\linewidth]{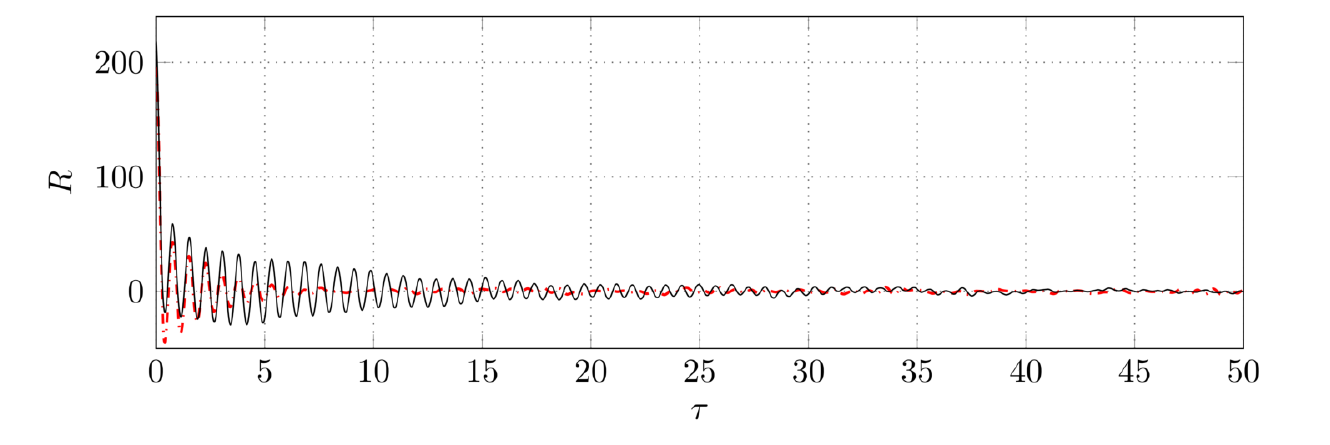}
	\caption{Autocorrelation function 
                 of the Lorenz system (black solid line) and the  cluster-based network model (red dashed line).}
	\label{fig:CorreFunctionLorenz}
\end{figure}

Figure \ref{fig:ProbLorenzCNM_Data} compares the asymptotic population
$\bm{p}^\infty$ predicted by CNM with the population from a long-term simulation.
The CNM statistics are based on $20,000$ transitions 
while the integration of the Lorenz equations is performed over 5000 time units.
Both statistics correspond to roughly 800 periods 
found to be sufficient for an accurate statistics.
The relative error of the CNM is no more than approximately 10\%.
This error does not decrease with much larger integration times,
but is linked to the coarse-graining of the state to Voronoi cells around the centroids.
The assumed constant transition time $\tau_{ij}$ for all trajectories 
from cluster $j$ to cluster $i$ is a crude assumption.
In fact, the transition times can vary by a large factor
and can thus give rise to significant systematic errors.
A more accurate transition model may, for instance, 
include earlier transitions for a more realistic representation of the trajectory.
Intriguingly, the CMM has an error of only 0.5\% 
which is one order of magnitude lower.
Due to the stroboscopic monitoring of the CMM states, 
no estimates of the transition times are required
and one source of systematic errors is excluded by construction.

A distinguishing feature of the CNM is the resolution of the temporal dynamics
illustrated in figure \ref{fig:ModeEvolution}.
The evolution of the model-based trajectory 
is hardly distinguishable 
from the one obtained by numerical integration.
Smoothness of the CNM trajectory has been achieved by splines 
connecting the states between two consecutive centroid visits.
Yet, the oscillatory amplitude growth in both wings 
cannot be resolved with this low cluster-based resolution.
The CNM can only follow the simulations for a short time period,
as nearby trajectories exponentially diverge with Lyapunov exponent 2.16 \citep{Wolf1985p}
and the initial separation in each cluster is already large.
Yet, the fluctuation amplitude, frequency content and bi-modality is well reproduced.
A CNM with $K=100$ clusters yields more realistic dynamics
but require orders of magnitude more simulation data.
On the least-order extreme, 
a CNM with 2 or 3 clusters only coarsely 
resolves the transitions between both ears of the Lorenz attractor,
not the growing oscillations in each ear.

Finally, the autocorrelation of the simulation (black solid curve)
and the CNM (red dashed curve) is presented for aggregate comparison 
in figure \ref{fig:CorreFunctionLorenz}.
CNM roughly reproduces the fast oscillatory decay 
of the autocorrelation function in the first five periods.

\section{Cluster-based reduced-order modelling of the mixing layer}
\label{ToC:ML}

In this section, the cluster-based models are applied 
to a two-dimensional incompressible mixing layer 
with Kelvin-Helmholtz vortices undergoing vortex pairing. 
The flow configuration of the mixing layer 
and the employed direct Navier-Stokes solver is presented in \S~\ref{S2:MLsimulation}. 
In \S~\ref{S2:FlowFeatures},
the dominant flow features of the mixing layer are presented.
Then (\S~\ref{ToC:ML:Clustering}), 
the snapshots of 
incompressible mixing layer are coarse-grained into centroids.
Following \citet{Kaiser2014jfm}, 
a cluster-based Markov model (CMM,  \S~\ref{ToC:ML:MarkovModel})
is developed as benchmark for the proposed network model (CNM,  \S~\ref{ToC:ML:NetworkModel}). 

\subsection{Flow configuration and direct numerical simulation}
\label{S2:MLsimulation}

The two-dimensional incompressible mixing layer 
and a velocity ratio of 3:1 
is considered as the test plant in this paper. 
The velocity ratio is a common choice 
in the literature \citep{Comte1998ejmb,Noack2005jfm,Kaiser2014jfm}.
The high- and low-speed streams have velocities $U_1$ and $U_2$, respectively.
The convection velocity $U_c$ 
of coherent structure is well approximated
by the average velocity \citep{Monkewitz1988jfm}:
\begin{equation}
\label{convectivevelocity}
U_{c}=\frac{U_{1}+U_{2}}{2}.
\end{equation}
The initial vorticity thickness is denoted by $\delta_0$.
The Newtonian fluid is characterized 
by the density $\rho$ and kinematic viscosity $\nu$.
The flow characteristics are described 
by the Reynolds number based on the convection velocity
$Re = U_c \delta_0 / \nu$ and velocity ratio.
In the sequel, 
 all quantities are assumed to be non-dimensionalized
with the initial vorticity thickness $\delta_0$,
the high-speed velocity $U_1$
and the density $\rho$.
The Reynolds number is set to 200.

The flow is described in a Cartesian coordinate system $(x,y)$
with the origin at maximum gradient location of the inlet profile.
The $x$-axis points in the streamwise direction and 
the $y$-axis points in the direction of the high-speed stream.
The velocity components  in $x$- and $y$-direction
are denoted by $u$ and $v$ respectively.

\begin{figure}
	\centering
	\def\svgwidth{0.8\linewidth}
	\includegraphics[width=\linewidth]{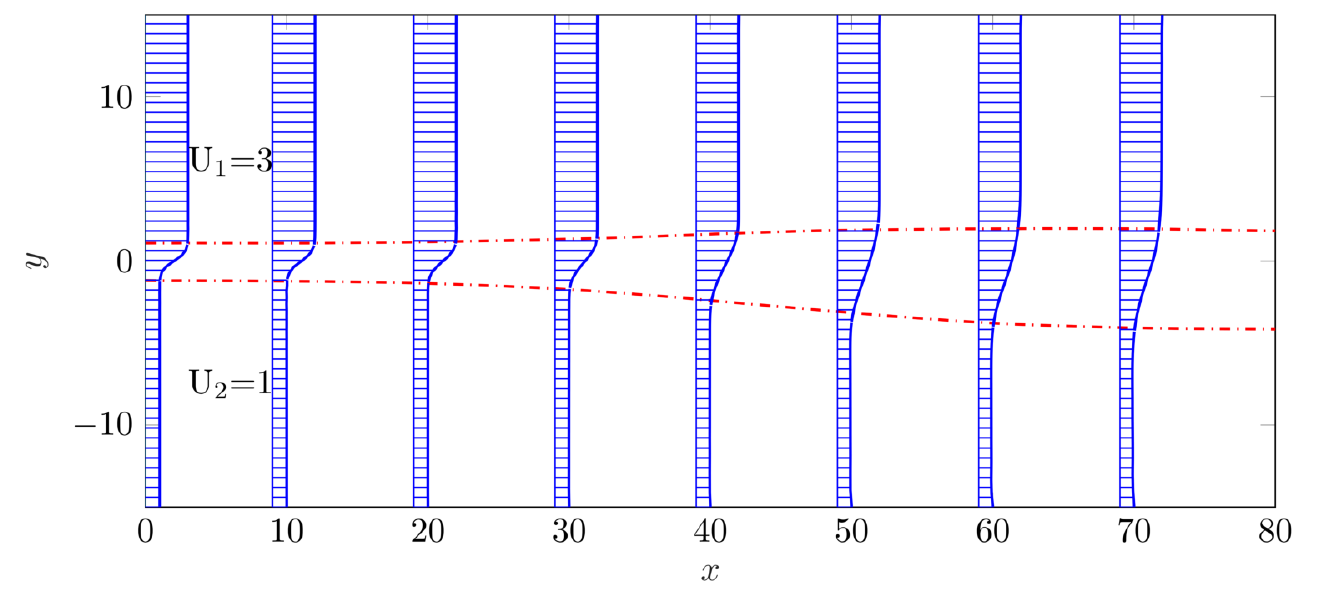}
	\caption{Numerical simulation sketch of incompressible mixing layer. 
		An unperturbed tanh velocity profile
		$u(y) =2+\tanh(2y)$ is patched in the inlet of a rectangular domain.
		Time-averaged streamwise velocity profiles separated by $\Delta x $=10 are 
		visualized by blue lines. 
		The red dashed curves mark the mixing layer thickness,
                We chose the 90\% thickness of the profile starting with the average velocity, 
                i.e., $u=2.9$ and $u=1.1$.}
	\label{Fig:SimulationSketch}
\end{figure}

Figure \ref{Fig:SimulationSketch} illustrates the
rectangular computational domain 
\begin{equation}
\Omega : = \{ ( x , y ) \in {\cal R}^2 : 0 \leq x \leq 80 \> \wedge \> \vert y \vert \leq 15 \}
\end{equation}
with 10237 nodes and 2248 triangular elements.
The location vector is denoted by $\boldsymbol{x}=(x,y)$.
Similarly, the velocity vector is denoted by $\boldsymbol{u}=(u,v)$. 
The inlet velocity profile reads
\begin{equation}
\label{Eqn:InletProfile}
u=2+\tanh \left( \frac{2y}{\delta_{0}} \right), 
\quad v=0, \quad \hbox{where} \quad \delta_{0}=1.
\end{equation}
The Kelvin-Helmholtz vortices are triggered at the inlet
by a  stochastic perturbation of the $u$-component for $y \in [-2,2]$ 
with  a standard deviation of $0.01 U_c$.

The streamwise extent of the domain is 80.
This corresponds to a downwash time of 40 given the convection velocity of 2.
This is the minimum time for the transient time
as all initial interior vortices will leave the domain.
A simulation over 400 convective units corresponds to
10 downwash times.
This period is found to be sufficient 
for a good statistics of the mean value and fluctuation level.
One simulation yields $M=10,000$ velocity snapshots 
$\boldsymbol{u}^{m}(\boldsymbol{x})=\boldsymbol{u}(\boldsymbol{x},t^m)$,
where the sampling times  $t^m = 0.04 \> m$ start 
with $t=0$ in the converged post-transient phase.
The sampling frequency $25$ is two orders of magnitude larger
than the dominant shear-layer frequency of $f=0.1075$
in the most active downstream region.

An in-house direct numerical simulation solver 
was employed to simulate the incompressible mixing layer. 
This solver is based on the Finite-Element Method (FEM)
with third-order Taylor-Hood elements 
with implicit third-order time integration.
The solver has been used for numerous configurations,
like the cylinder wake \citep{Noack2016jfm}, 
the mixing layer \citep{Shaqarin2018fdr},
the fluidic pinball \citep{Ishar2019jfm}, to name only a few.

\begin{figure}[htb]
	\centering
	\sidecaption{subfig:a}
	\raisebox{-\height}{\def\svgwidth{1\linewidth}\includegraphics[width=\linewidth]{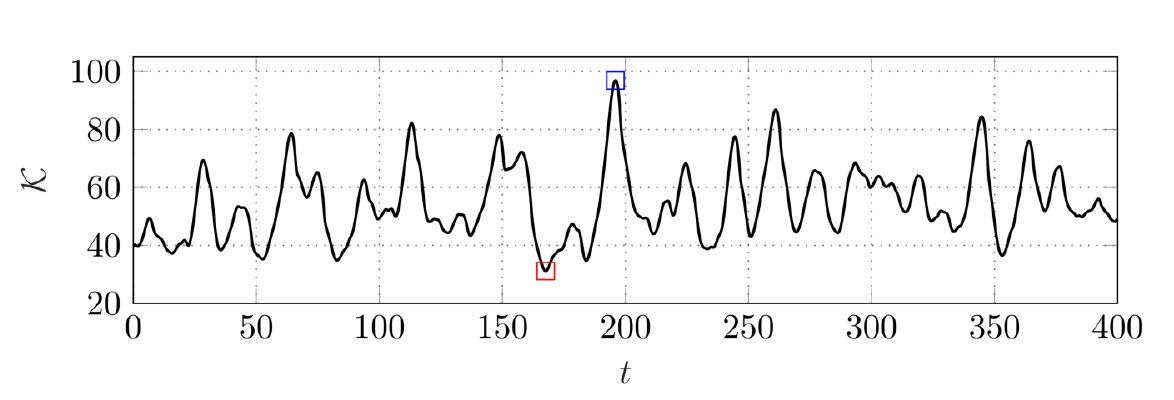}}%
	\par
	\sidecaption{subfig:b}
	\raisebox{-\height}{\def\svgwidth{0.99\linewidth}\includegraphics[width=\linewidth]{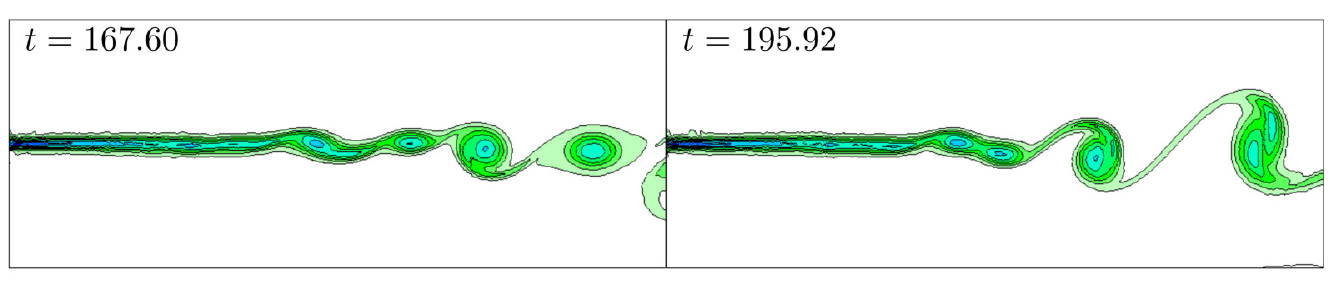}}
	\caption{Mixing layer simulation. 
                 \subref{subfig:a} Energy fluctuation over time with the maximum marked by a blue square and minimum by a red square.
		 \subref{subfig:b} Vorticity fields associated with the maximum and minimum fluctuation energy.
The minimum (left) corresponds to a K-H vortex at $t=167.60$, 
The maximum (right) features vortex pairing  at $t=195.92$.
The curves represent the isolines of vorticity.
Higher values corresponds to darker green areas.}
		\label{Fig:VorticitySnapshots}
\end{figure}

\begin{figure}[htb]
	\centering
	\def\svgwidth{0.8\linewidth}
	\includegraphics[width=0.8\linewidth]{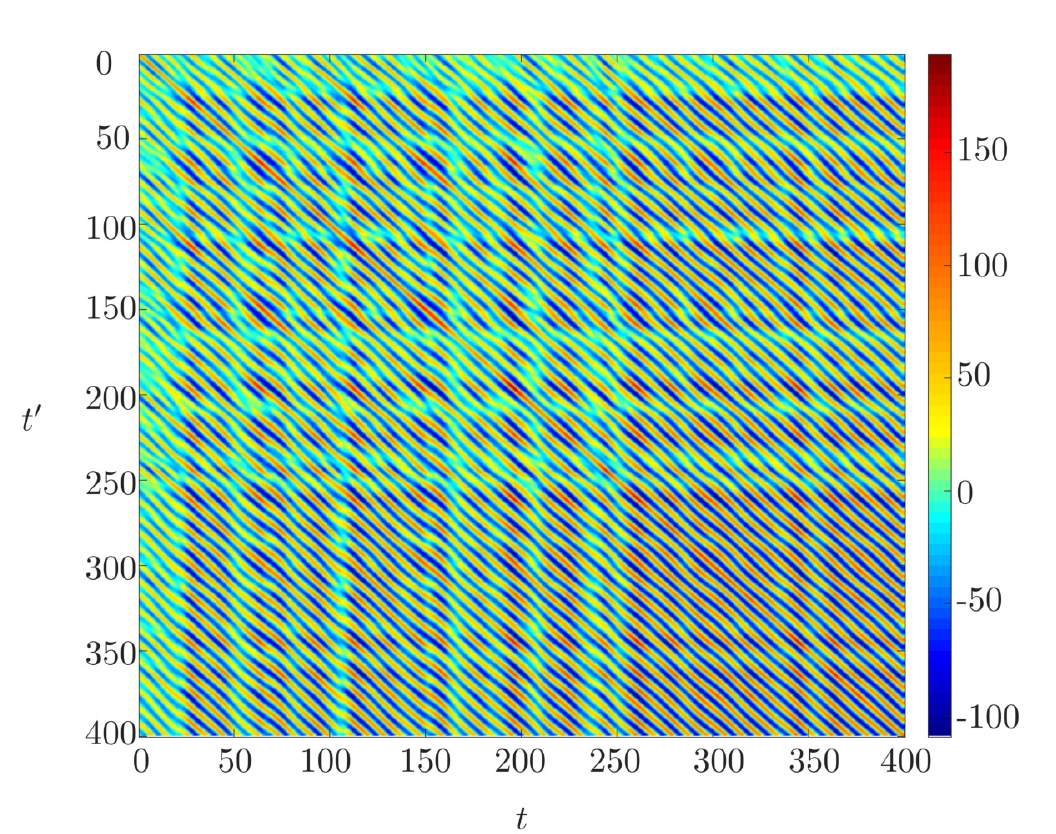}
	\caption{Autocorrelation matrix \eqref{Eqn:CorrelationMatrix} 
                 of the mixing layer for $t ,t^\prime \in [0,400]$.
                 The value is presented in the colorbar.
                 The plot is based on  401 snapshots  collected at uniform time steps $\Delta t=1$.}
		\label{Fig:CorrelationMatrix}
\end{figure}
\subsection{Flow features}
\label{S2:FlowFeatures}

The incompressible mixing layer 
exhibits two typical behaviours.
First, the initial dynamics is characterized 
by the roll-up of vorticity
originating from  the Kelvin-Helmholtz (K-H) instability 
(see left in figure \ref{Fig:VorticitySnapshots}\subref{subfig:b}).
Second, these vortices pair further downstream 
as can bee seen at the outlet region of figure \ref{Fig:VorticitySnapshots}\subref{subfig:b}(right).
This vortex pairing contributes to the mixing layer growth.
The location of vortex pairing may change significantly in time.
Upstream  (downstream) vortex pairing is associated
with high (low) fluctuation energy.

The time-averaged velocity field 
in figure \ref{Fig:SimulationSketch} shows the
mixing layer growth. 
The velocity thickness is visualized by a red dashed line 
and is defined as the distance between transverse locations where the mean streamwise
velocity was equal to $U_{1}-0.05 \Delta U$ and $U_{2}+0.05 \Delta U$. 
The mixing layer  thickness increases significantly between $x=30$ and $x=60$.
Here, vortex pairing leads to this thickness increase.

The temporal dynamics may be inferred from the evolution of the fluctuation energy
in figure \ref{Fig:VorticitySnapshots}\subref{subfig:a}.
The fluctuations indicate narrow bandwidth oscillatory behaviour.
More refined insights may be gained from the correlation function
between the flows at time $t$ and $t^\prime$,
\begin{equation}
C(t,t^{\prime}) = \int\limits_{\Omega}  \bm{u}^\prime \left ( \bm{x},t         \right ) \cdot 
                                                         \bm{u}^\prime \left ( \bm{x},t^\prime  \right )
                                                         \>  d\bm{x}
\label{Eqn:CorrelationMatrix}
\end{equation}
Figure \ref{Fig:CorrelationMatrix} illustrates the autocorrelation matrix
for $t,t^\prime \in [0,400]$.
The fluctuation energy of figure \ref{Fig:VorticitySnapshots}\subref{subfig:a}
is quantified in the diagonal, ${\cal K}(t) = C(t,t)/2$. 
The wavy pattern indicates oscillatory coherent structures.
The changes from pure periodicity are caused by vortex pairing at a large range of streamwise locations.

\subsection{Clustering}
\label{ToC:ML:Clustering}
\begin{figure}
	\centering
	\def\svgwidth{1\linewidth}
	\includegraphics[width=\linewidth]{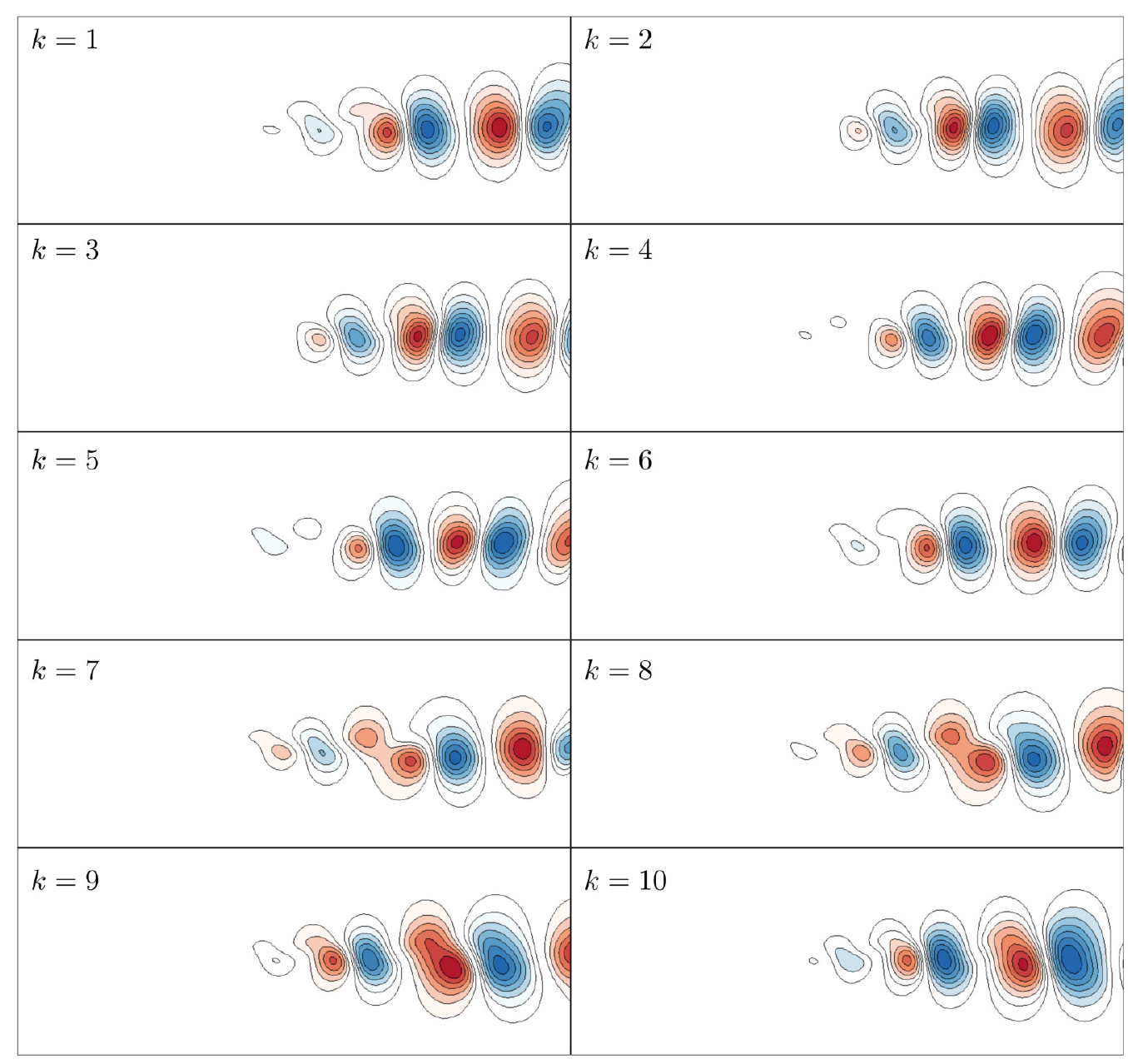}
	\centering
	\caption{The cluster centroids $\bm{c}_{k}$, $k=1, \ldots, 10$, of the mixing layer.
                 The transverse velocity fluctuation is depicted with contour lines. 
                 Red and blue regions mark positive and negative values.}
	\label{fig:Centroids}
\end{figure}
Both considered reduced-order models are based on the direct numerical simulation of 
the two-dimensional incompressible mixing layer described in \S~\ref{S2:MLsimulation}.
$M=10,000$ velocity field snapshots of the post-transient phase 
are sampled with a time step  $\Delta t = 0.04$.

The computational load of clustering 
is significantly reduced by an effectively lossless POD compression 
detailed in Appendix \ref {ToC:Method:POD}. In fact, all 
operations are performed on the POD amplitude vector
$\bm{a}= [a_{1},a_{2},...,a_{N}]^T$ instead of the snapshots. 

The $M$ snapshots are clustered 
with the k-means++ algorithm into $K=10$ centroids.
This number is small enough to allow for the physical interpretation of all centroids
and all transitions but large enough for a meaningful reduced-order model.
Figure \ref{fig:Centroids} illustrates the transverse velocity fluctuation of the centroids.
The first six centroids show the streamwise convection of Kelvin-Helmholtz (K-H) vortices,
while the next four centroids resolve a vortex pairing (VP) event.
In centroid $7$, two vortices merge at the beginning of the vortex chain.
In the following three centroids, 
the merging is completed and leads to a large vortex.
Note that the VP centroids $k=7,8,9,10$ 
have pronounced vortices at a similar position as 
the KH centroids $k=4,5,6$, respectively. 
The structures of the KH and VP centroids are noticeably different.
The main vortices of the KH centroids are elliptical and the major axis is rotated in clockwise direction, i.e., 
the upper part of the vortices follow the faster stream.
In contrast, the main elliptical vortices of VP centroids are rotated in mathematically positive direction, i.e., 
the upper part of these vortices move upstream with respect to their center.

\begin{figure}
	\centering
	\sidecaption{subfig:a}
	\raisebox{-\height}{\def\svgwidth{0.3\linewidth}\includegraphics[width=0.44\linewidth]{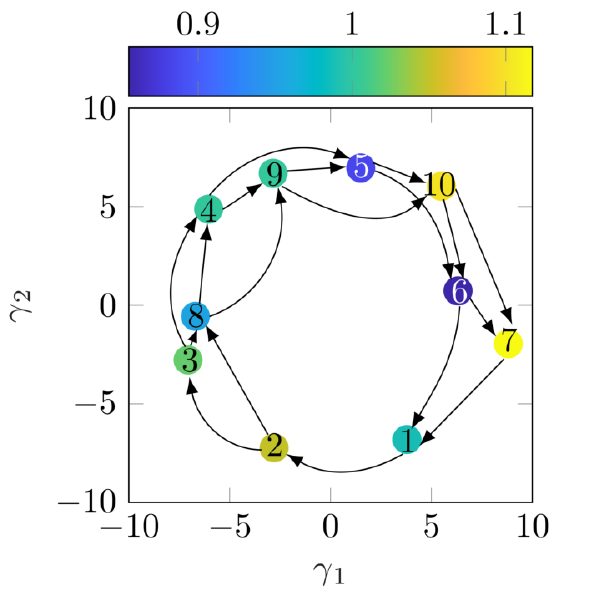}}%
	\hfill
	\sidecaption{subfig:b}
        \raisebox{-\height}{\def\svgwidth{0.37\linewidth}\includegraphics[width=0.44\linewidth]{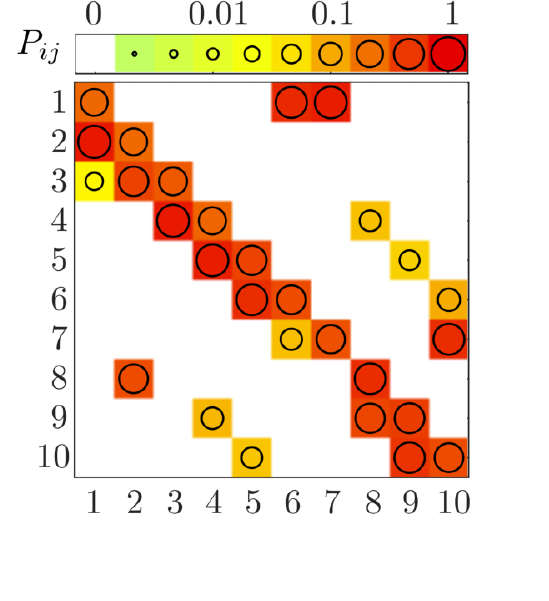}}
	\caption{Cluster-based Markov model of the mixing layer.  
		\subref{subfig:a}
		Proximity map of centroids. Each centroid is marked by a solid coloured circle. 
                The color denotes the relative energy content (see colorbar on top). 
                Unity corresponds to the average value.
               \subref{subfig:b}
                Transition matrix.
		The probability value is displayed by the background color 
		and the radius of responding circle.}
	\label{fig:DynamicsMarkov}
\end{figure}
The centroids represent characteristic stages in the mixing layer dynamics
as can be elucidated in a proximity map.
This map reflects the configuration matrix 
${\sf{\bm{D}}} = (D_{ij}) \in {\cal R}^{K \times K}$ 
comprising the distance between two centroids:
\begin{equation}
D_{i j}^{c} :=\left\|\bm{c}_{i}-\bm{c}_{j}\right\|_{\Omega},i,j=1,2...,K.
\label{Eqn:Distance}
\end{equation}
Following \citet{Kaiser2014jfm},
the proximity map is used to represent the configuration matrix ${\sf{\bm{D}}}$
in a two-dimensional feature space $\bm{\gamma} \in {\cal R}^2$ 
optimally preserving the relative distances.
The proximity map employs \emph{Classical Multidimensional Scaling} (CMDS) \citep{Mardia1979book}.
Figure \ref{fig:DynamicsMarkov}\subref{subfig:a} displays 
centroids close to a circle which is characteristic for vortex shedding.

\subsection{Markov model}
\label{ToC:ML:MarkovModel}

The temporal mixing-layer evolution
is characterized by the cluster transition matrix $\sf{\bm{P}}$ 
illustrated in figure \ref{fig:DynamicsMarkov}\subref{subfig:b}.  
$P_{ij}$ represents the probability of moving from cluster $j$ to $i$ in one forward time step.
Here, we choose a time step  $\Delta t^c=T/10=1$ 
where the $T=10$ is the dominant period of the evolved mixing layer. 

The cluster transition matrix reveals two cyclic groups.
The first group $1 \to 2 \to 3 \to 4 \to 5 \to 6 \to 7 \to 1$
is consistent with the convection process of the K-H vortex shedding
observed in the centroid visualization.
This periodic process corresponds 
to a nearly uniform clockwise rotation in the proximity map.
The second cyclic group $8 \to 9 \to 10 \to 7 \to 1 \to 2\to 8$
comprises VP centroids $k=8,9,10$ 
and shares two centroids with the K-H regime.
These dynamics also lead to a nearly uniform clockwise rotation in the feature space.
There are also transitions from the VP  to K-H regime,
e.g.\ $8 \to 4$, $9 \to 5$, $10 \to 6$ and $10 \to 7$
and in the opposite direction.
All these transitions are between similar centroids of both groups.
From the cluster index the orientation
of the main elliptical vortices can be inferred.
For $k \le 6$ ($k\ge 7$), 
the upper part of the vortices are displaced in (against) the direction of the flow
with respect to their centers.

\begin{figure}
	\centering
	\def\svgwidth{0.8\linewidth}
	\includegraphics[width=0.8\linewidth]{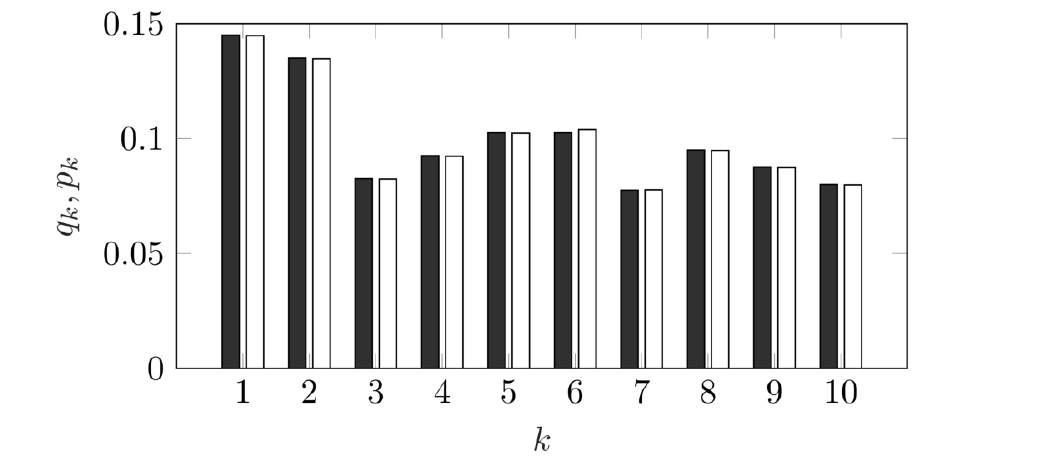}
	\caption{Cluster probability distribution of the mixing layer from the DNS data and the cluster-based Markov model (CMM).
                Solid rectangles denote the  probability $q_{k}$ from the DNS data.
                Open rectangles represent asymptotic values from the CMM after $l=35$ iterations.}
	\label{fig:ProbData_CMM}
\end{figure}
\begin{figure}[htb]
	\centering
	\def\svgwidth{0.8\linewidth}
	\includegraphics[width=1\linewidth]{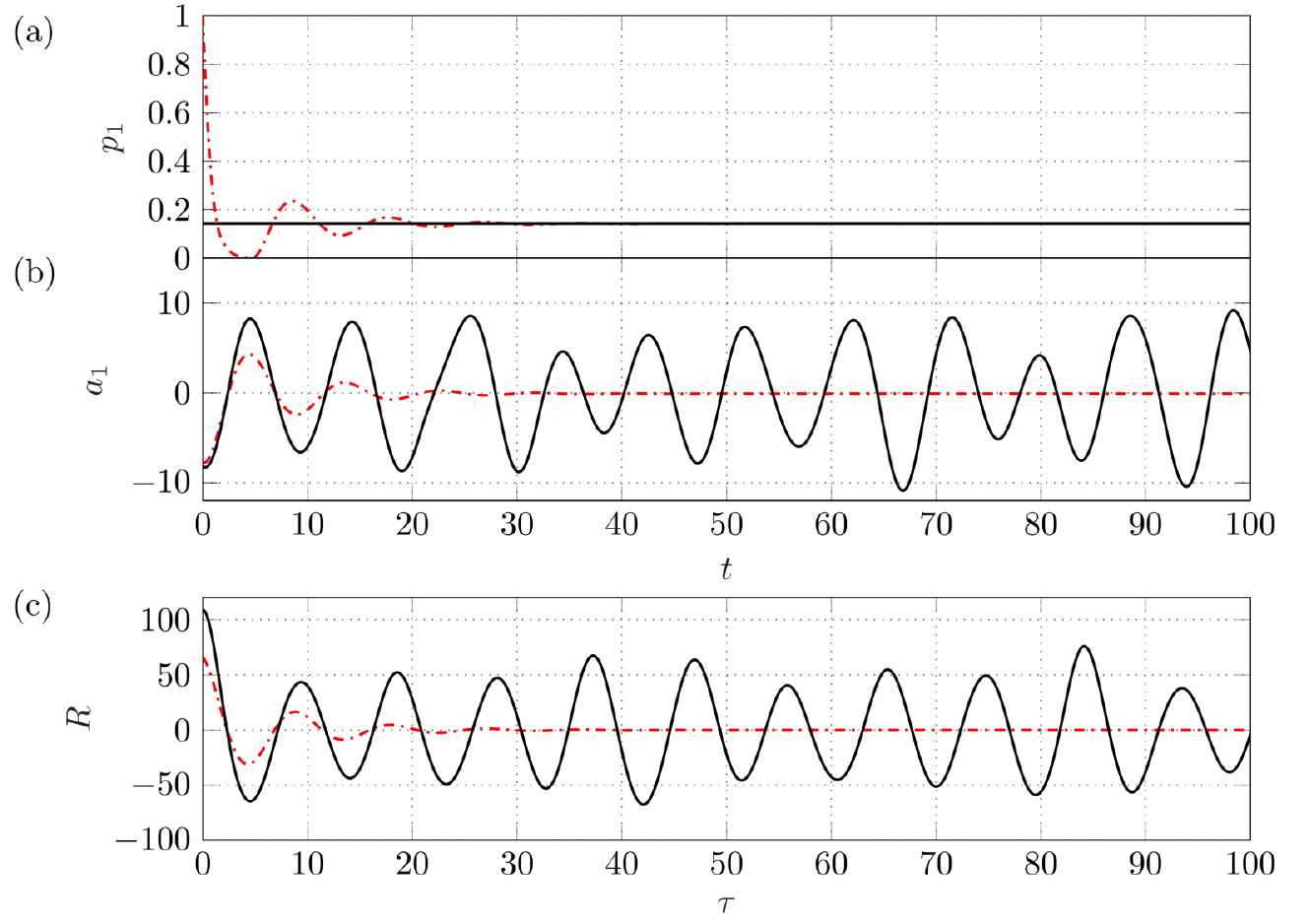}
	\caption{Dynamics of the mixing layer from DNS and the cluster-based Markov model (CMM). 
                (a) Probability evolution for  DNS (black solid line) and for CMM (red dashed line). 
                The probability of the first cluster $p_{1}$ 
                quickly converges to $p^\infty_i$ around $t=35$.
                The corresponding DNS value  is $0.1325$ and represented by a horizontal line. 
                (b) 
		The evolution of the first POD mode amplitude $a_{1}$ for DNS (black solid line) and CMM (red dashed line). 
		(c) The autocorrelation function for DNS (black solid line) and for CMM (red dashed line).}
	\label{fig:Evolution}
\end{figure}
The evolution of the cluster population vector $\bm{p}^l$ at $t=l \Delta t^c$
is investigated  by iterating  equation \eqref{Eqn:CMM}.
Figure \ref{fig:ProbData_CMM} compares the probability distribution 
of DNS data and the model-based asymptotic  vector $\bm{p}^{\infty}$. 
The agreement is astonishingly good for such a low-order model.
The probability vector converges quickly to a unique, stationary probability distribution near $t=20$.  

In figure \ref{fig:Evolution},
the dynamics of CMM is illustrated
for the first cluster probability $p_1$ 
and the first POD mode amplitude $a_1$ inferred from the flow state \eqref{Eqn:EvolutionVelocity}.
Starting point is direct numerical simulation starting at $t=0$ close to the first cluster $\bm{c}_1$
which corresponds to  the probability vector $\bm{p}=[1,0,0,0,0,0,0,0,0,0]^T$. 
The probability and POD mode amplitude of CMM 
show a convergence after around $l=35$ iterations or, equivalently, $t\approx 35$.
The solid horizon line denotes $q_1$, i.e.\ the population of the first cluster from DNS data. 
The POD mode amplitude $a_1$ performs three strongly damped oscillations before vanishing. 

Figure \ref{fig:Evolution}(c) shows 
an oscillating quickly decaying autocorrelation function of CMM 
which is consistent with the observations for $a_1$ and $p_1$.
In contrast, the autocorrelation function associated with the DNS
keeps oscillating around with an amplitude around 50\% of the average fluctuation level.
This level indicates that half of the fluctuation energy resides 
in repeating oscillatory flow structures
while the other half is of non-repeating stochastic nature.

\subsection{Network model}
\label{ToC:ML:NetworkModel}

\begin{figure}
	\centering
	\sidecaption{subfig:a}
	\raisebox{-\height}{\def\svgwidth{0.445\linewidth}
	\includegraphics[trim=0 0 0 0,clip,width=0.44\linewidth]{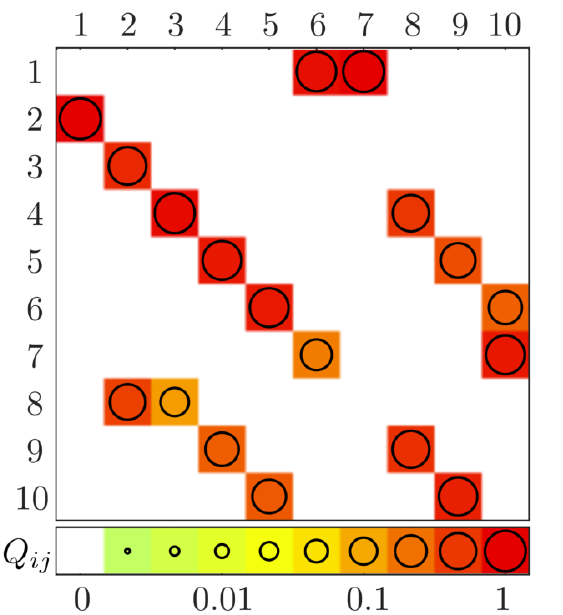}}
	\hfill
	\sidecaption{subfig:b}
	\raisebox{-\height}{\def\svgwidth{0.445\linewidth}
	\includegraphics[trim=0 0 0 0,clip,width=0.44\linewidth]{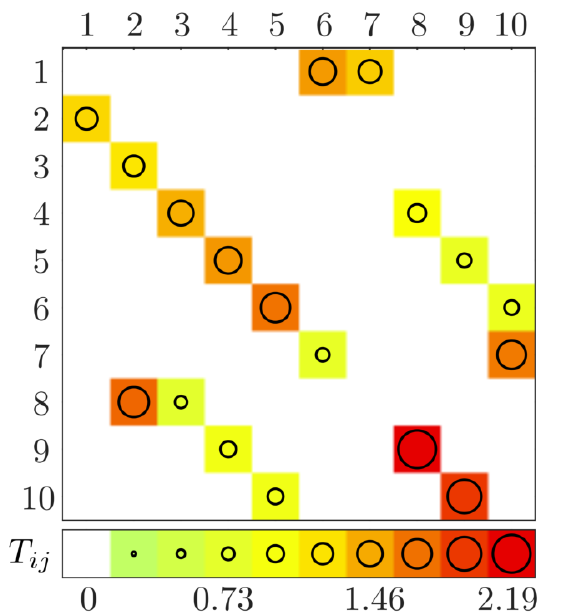}}
	\caption{Dynamics of the cluster-based network model for the 
		mixing layer.
                \subref{subfig:a} Direct transition matrix.  
                \subref{subfig:b} Averaged transition time. 
                The non-vanishing values are denoted by the circle radius 
                and the color code from the bottom caption.
		\label{fig:DirectTransitionMatrix}}
\end{figure}
In this section, 
a \emph{Cluster-based Network Model} (CNM) 
is developed using the same snapshot data and same centroids.
Starting point for the dynamic network 
is the cluster affiliation function $k(t)$. 
Following \S~\ref{ToC:Method:NetworkModel},
the direct cluster transition matrix $\sf{\bm{Q}}$ 
with associated average transition times ${\sf{\bm{T}}}$ are derived.
Figure \ref{fig:DirectTransitionMatrix} illustrates both matrices.
These matrices have the almost same structure as the Markov model
except for the diagonal elements which are vanishing by design. 
In other words,
\begin{subequations}
\begin{eqnarray}
Q_{ii}&=&T_{ii}=0 \quad \forall i \in \{1,\ldots,K\}    ,
\label{Eqn:StructureCNMa}
\\ H[P_{ij}]&=&H[Q_{ij}]= H[T_{ij}] \quad \forall i, j \in \{1,\ldots,K\}  \wedge i \not = j ,
\label{Eqn:StructureCNMb}
\end{eqnarray}
\label{Eqn:StructureCNM}
\end{subequations}
$H$ being again the Heaviside function.
Vanishing diagonal elements \eqref{Eqn:StructureCNMa} arise from the requirement of non-trivial transitions.
Theoretically, the trajectory may terminate in a cluster, 
like in a stable fixed point of a linear dynamical system.
This case is not compatible with the goal to model a well-resolved non-trivial attractor and shall be ignored in this study.
Equation \eqref{Eqn:StructureCNMb} requires a sufficiently small time-step of the CMM.
Otherwise, the stroboscopic view on the trajectory may miss a crossing of an intermediate cluster. 
This happens with the transition from $1 \to 2 \to 3$ in one CMM time step $\Delta t^c$.
Hence, $P_{31} \not = 0$ while $Q_{31} = 0$.
However, this is a rare event as indicated by the small value of $Q_{31}$.

An inspection of ${\sf{\bm{T}}}$ reveals that the transition time between K-H and VP centroids is relatively small. 
This is consistent with the closeness of the corresponding centroids in the proximity map  (figure \ref{fig:DynamicsMarkov}\subref{subfig:a}).
An exception is the transition between K-H centroid 2 to VP centroid 8 which are well separated in the proximity map.
Intriguingly, 
the transitions within the K-H and VP regime are also strongly correlated with the distances depicted in the proximity map.
For instance, the smallest (largest)  inner-regime transition from centroid $6$ to $7$ ($8$ to $9$) 
is associated with a small (large) distance in the proximity map.
The physical interpretation of the cycle-to-cycle variations
of the CMM persist for CNM.

\begin{figure}
	\centering
	\def\svgwidth{0.8\linewidth}
	\includegraphics[width=0.8\linewidth]{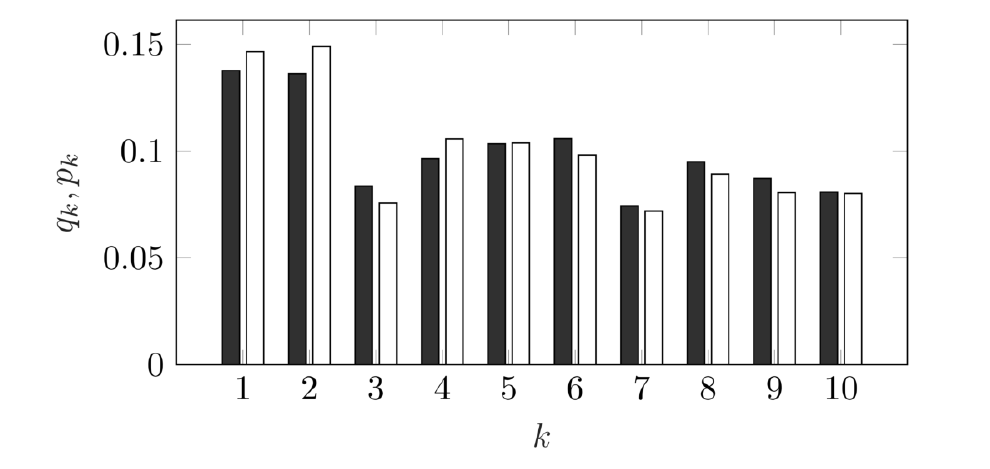}
	\caption{Cluster probability distribution of the mixing layer 
                from DNS (solid rectangle) and cluster-based network model  (open rectangle).
                The modelled values are obtained from simulating  $20,000$ clusters transitions.}
	\label{fig:ProbCNM_Data}
\end{figure}
In the following, the temporal dynamics of CNM is investigated
based on the identified centroids $\bm{c}_{k}$,
the description of their connectivity DTM $\sf{\bm{Q}}$,
and their flight times ${\sf{\bm{T}}}$.
Like a POD model,  CNM is a grey-box model 
resolving the temporal dynamics and the associated coherent structures.
We choose  cluster $k=1$ as initial condition for DNS and for the CNM
and integrate over $l=20,000$ transitions.
In figure \ref {fig:ProbCNM_Data},
the asymptotic cluster population $\bm{p}^\infty$ from equation \eqref{Eqn:CNM:pi}
is compared with $\bm{q}$ from the DNS.
The discrepancies of few percent seem 
expectable and tolerable for a 10-cluster model.
This difference is not cured by increasing the amount of transition data in CNM.
Intriguingly, the probability distribution of the CMM 
displayed in figure \ref{fig:ProbData_CMM} is significantly more accurate.
This behaviour can be linked to the simple transition time estimate 
which employs one single average value for a large range of observed transition times.
We have developed more refined and more accurate transition time estimates 
leading to much better agreements of the cluster probability distributions. 
The price is increased complexity of the CNM
which we deemed not helpful for our first publication.
  
\begin{figure}
	\centering
	\def\svgwidth{0.8\linewidth}
	\includegraphics[width=\linewidth]{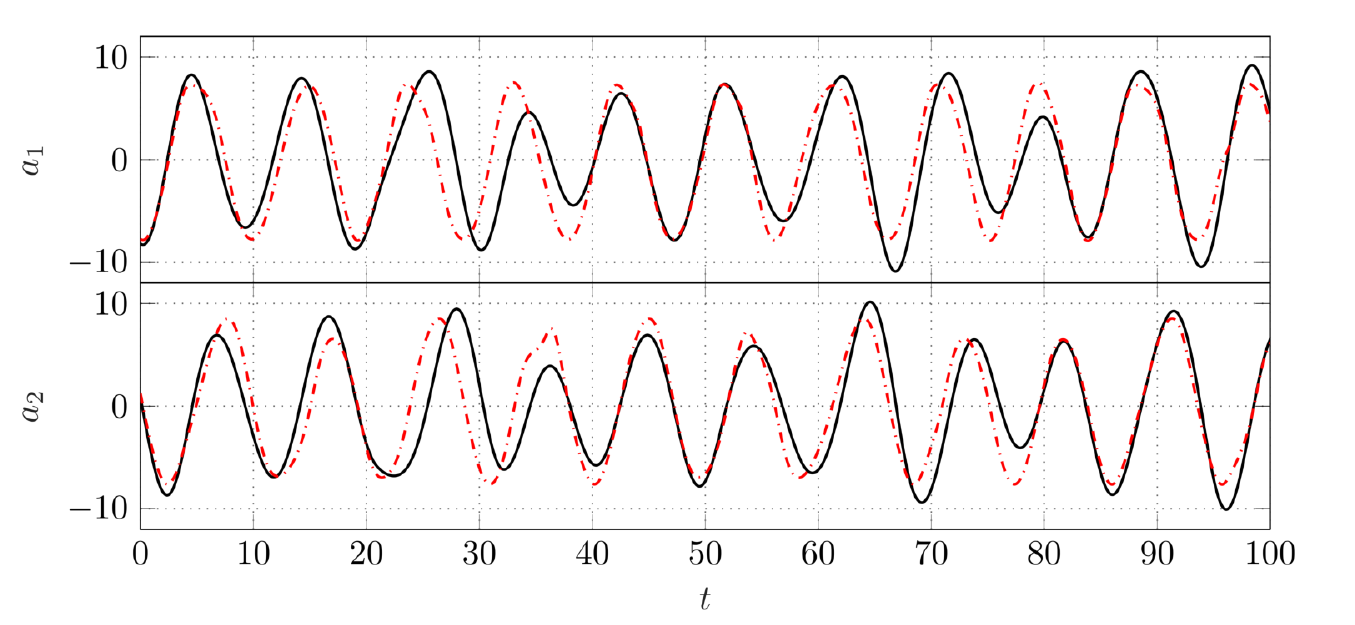}
	\caption{Evolution of mode amplitudes $a_{1}$, $a_{2}$ for the mixing layer $t\in[0,100]$. 
                 The curves correspond to DNS (black solid line) and cluster-based network model (red dashed line).}
	\label{fig:ModeEvol}
\end{figure}
Figure \ref{fig:ModeEvol} shows the evolution of the first two POD mode amplitudes (red dashed curve).
The CNM tracks well the amplitude and phase of the DNS over $100$ time units.
Like for the Lorenz system, the temporal evolution is smoothed by a spline
and does not use the non-smooth uniform motion between two consecutive centroid visits. 

\begin{figure}[htb]
	\centering
	\def\svgwidth{0.3\linewidth}
	\includegraphics[trim=0 0 0 0 clip,width=\linewidth]{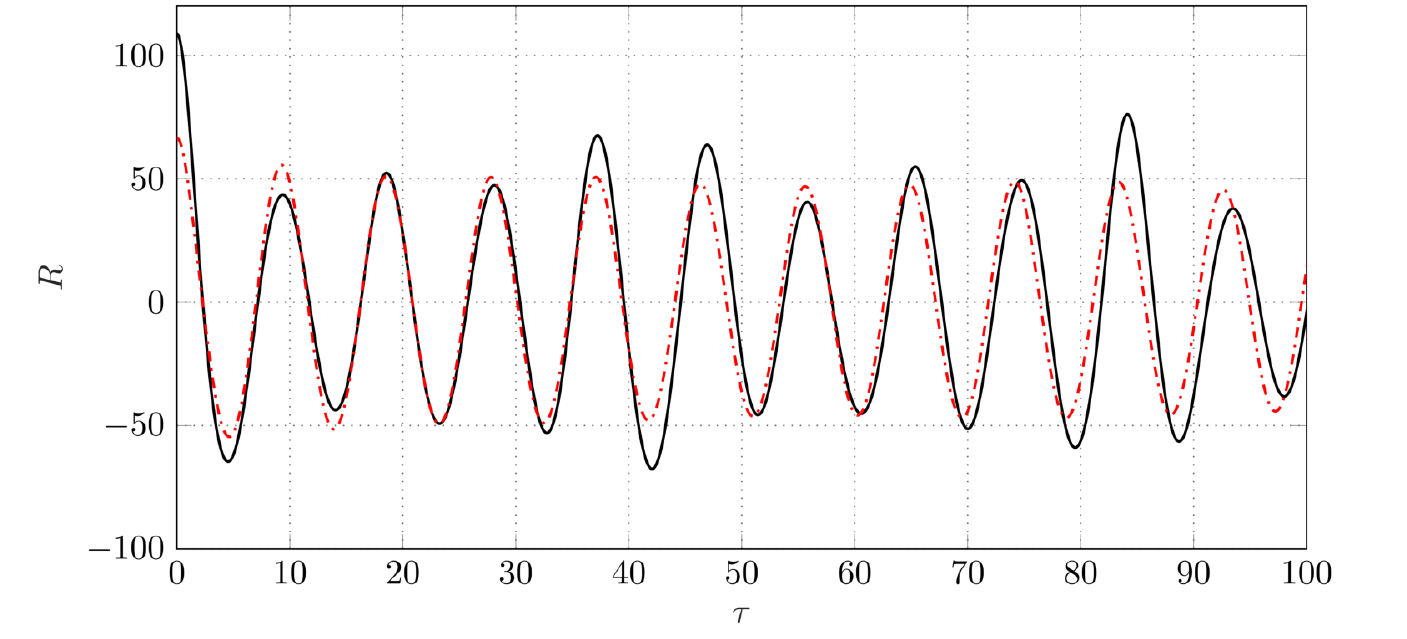}
	\caption{Autocorrelation function of the mixing layer 
                 for DNS (black solid line) 
		 and cluster-based network model (red dashed line).}
	\label{fig:CorreFunction}
\end{figure}

Figure \ref{fig:CorreFunction} compares the autocorrelation function of the CNM and the DNS.
We intentionally do not normalize this function to reveal the resolved fluctuation level
at vanishing time delay.
As expected, the model-based fluctuation level 
is significantly lower than the DNS value.
This  difference is quantified by the unresolved inner-cluster variance.
Intriguingly, CNM and DNS functions  become already similar after half a period.
The asymptotic fluctuation level represents coherent structures 
which are well resolved by the chosen centroids
and serve as coarse-grained recurrence points of the DNS.
Due to the dominant oscillatory dynamics,
 the autocorrelation does not vanish with increasing time.
The good reproduction autocorrelation function 
is a posteriori justification for the chosen cluster number.

\section{Cluster-based network modelling of the actuated turbulent boundary layer}
\label{ToC:TBL}
In this section, the cluster-based network modelling 
is implemented on a three-dimen\-sional actuated turbulent boundary layer.
First (\S\ \ref{ToC:Simulation:TBL}), 
the flow configuration and the large-eddy simulation is described.
The clustering results, 
which follow the same coarse-graining approach as for the shear layer,
are presented in \S~\ref{ToC:TBL:Clustering}.
A cluster-based network model 
is developed and assessed in \S~\ref{ToC:TBL:CNN}.

\subsection{Flow configuration and large-eddy simulation}
\label{ToC:Simulation:TBL}
In this section, the actuated turbulent boundary layer configuration for skin friction reduction is detailed.
In particular, the actuation mechanism is presented, and the numerical setup is described.
For more details, the reader is referred to \citet{Albers2019b} and \citet{Fernex2020}.
\begin{figure}
		\includegraphics[width=\linewidth]{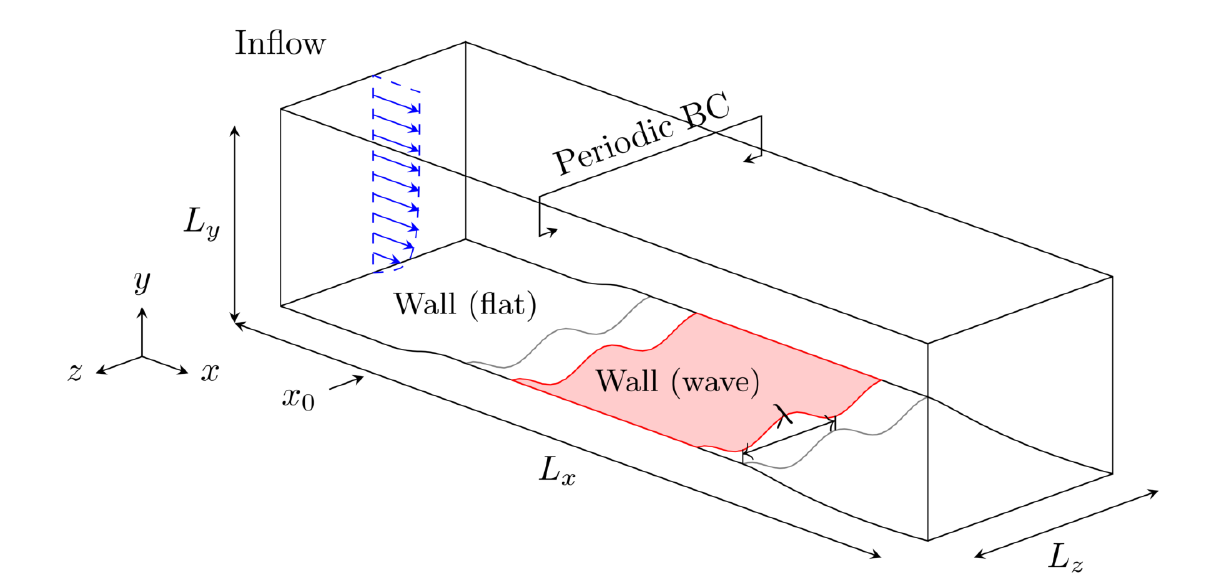}
		\caption{Overview of the physical domain of the actuated turbulent
			boundary layer flow, where $L_x, L_y,$ and $L_z$ are the domain
			dimensions in the Cartesian directions, $\lambda$ is
			the wavelength of the spanwise traveling wave, and $x_0$ marks the
			actuation onset. The shaded red surface $A_\mathrm{surf}$ marks the integration area of
			the wall-shear stress $\tau_w$.}
		\label{fig::grid}
\end{figure}
The fluid flow is described in a Cartesian frame of reference where
the streamwise, wall-normal, and spanwise coordinates are denoted by
$\bm{x} = (x,y,z)$ and the velocity components by $\bm{u} =
(u,v,w)$. The Mach
number is set to {\it Ma}$ = 0.1$ corresponding to a  nearly incompressible
flow. 
An illustration of the rectangular physical domain is
shown in figure \ref{fig::grid}. A momentum thickness of $\theta = 1$ at
$x_0$ is achieved such that the momentum thickness based Reynolds
number is $Re_\theta = 1000$ at $x_0$. The domain length and height in the
streamwise and wall-normal direction are $L_x = 190\,\theta$ and $L_y
= 105\,\theta$. In the spanwise direction, different domain widths $L_z
\in [21.65\,\theta, 108.25\,\theta]$ are used to simulate different
actuation wavelengths.

At the domain inlet, 
a synthetic turbulence generation method
is applied to generate a natural turbulent boundary layer flow 
after a transition length of 2-4 boundary layer thicknesses
\citep{Roidl2013}. 
Characteristic boundary conditions are used at the
domain exit and a no-slip wall boundary condition is enforced at
the lower domain boundary for the unactuated and actuated wall.

The actuation is performed 
by a transverse travelling wave on the surface.
The corresponding wall motion 
is prescribed 
by the space- and time-dependent function
\begin{equation}
y_{\text{wall}}^+(z^+,t^+) =
A^+ \cos\left( \frac{2\pi}{\lambda^+}z^+ - \frac{2\pi}{T^+}t^+ \right)
\label{Eqn:Actuation}
\end{equation}
in the interval $-5 \leq x/\theta \leq 140$.
The quantities $\lambda^+$, $T^+$, and $A^+$ denote the wavelength, period, and
amplitude in inner coordinates, i.e., the parameters are scaled by the viscosity
$\nu$ and the friction velocity of the unactuated reference case
$u^n_{\tau}$.
In the area just upstream
and downstream of the wave actuation region, a spatial transition is
used from a flat plate to an actuated plate and vice versa \citep{Albers2019b}.
In total, 38 actuation configurations with wavelength $\lambda^+ \in [200,500,3000]$, period
$T^+ \in [20,120]$, and amplitude $A^+ \in [10,78]$ are simulated.
In the current study, 
we  model one test case with $\lambda^+=1000$, $T^+=120$, and $A^+ =40$ 
which yields the largest drag reduction of 3\% found at that wavelength.
These actuation parameters correspond to case N36 in Table 3 of \citet{Ishar2019jfm}
and in Table 2 of \citet{Albers2020ftc}.

The physical domain is discretized by a structured block-type mesh
with a resolution of $\Delta x^+ = 12.0$ in the streamwise and $\Delta
z^+ = 4.0$ in the spanwise direction. In the wall-normal direction, a
resolution of $\Delta_y^+|_{\mathrm{wall}} = 1.0$ at the wall is used
with gradual coarsening away from the wall. 
Depending on the  domain width, the meshes consist of $24$ to $120$ million cells.

The actuated flat plate turbulent boundary layer flow is governed by
the unsteady compressible Navier-Stokes equations in the arbitrary
Lagrangian-Eulerian formulation for time-dependent domains. 
A second-order accurate finite-volume approximation of the governing
equations is used in which the convective fluxes are computed by the
advection upstream splitting method (AUSM) and time integration is
performed via a 5-stage Runge-Kutta scheme. 
The smallest dissipative scales are implicitly modelled through the numerical dissipation of the
AUSM scheme. This monotonically integrated large-eddy simulation
approach \citep{Boris1992fdr} is capable of accurately capturing all
physics of the resolved scales \citep{Meinke2002cf}.

The actuated simulations are
initialized by the solution from the unactuated reference case and
the temporal transition from the flat plate to the actuated wall is
initiated. 
When a converged state of the friction drag is obtained, statistics
are collected for $t U_\infty / \theta = 1250$ convective times.

\begin{figure}
	\centering
	\sidecaption{subfig:a}
	\raisebox{-\height}{\def\svgwidth{0.6\linewidth} \includegraphics[trim=0 0 0 0,clip,width=0.4\linewidth]{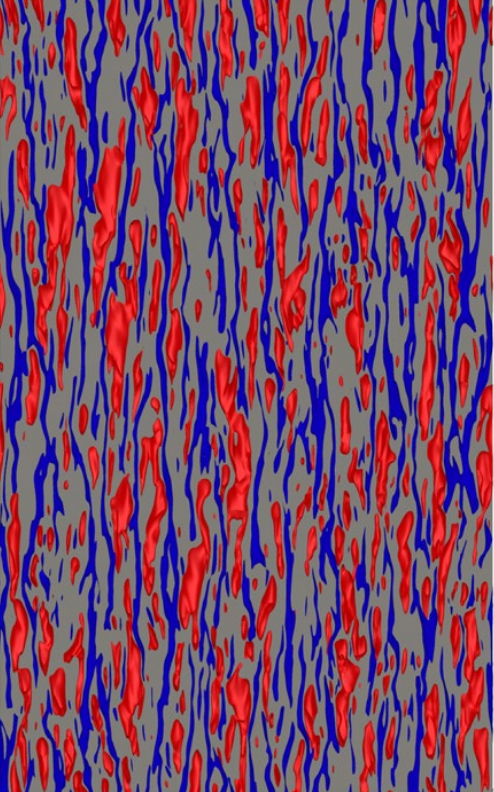}}%
	\hfill
	\sidecaption{subfig:b}
	\raisebox{-\height}{\def\svgwidth{0.6\linewidth} \includegraphics[trim=0 0 0 0,clip,width=0.4\linewidth]{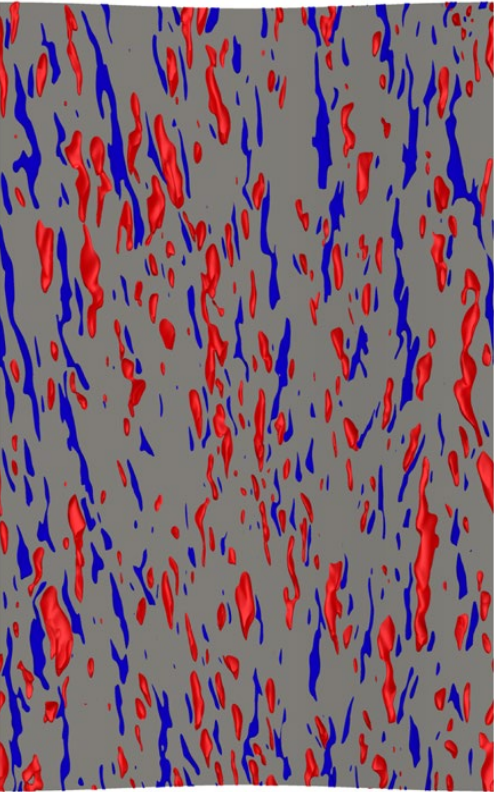}}
	\caption{Contours of the random streamwise velocity fluctuations in the near-wall region of \subref{subfig:a} 
	a non- actuated reference case and \subref{subfig:b} the actuated case. 
	The actuation strongly diminishes the near-wall streaks intensity. 
	The figure is from \citet{Albers2020ftc}.}
\label{fig:QFlow}
\end{figure}
The actuation effects on the near-wall  flow features are illustrated in figure \ref{fig:QFlow}, 
which shows contours of the streamwise velocity fluctuation of a reference 
natural (\ref{fig:QFlow}\subref{subfig:a}) and the actuated case (\ref{fig:QFlow}\subref{subfig:b}).
The intensity of the near-wall streaks, which are known to contribute to skin-friction, 
are observed to strongly diminish with the actuation.

\subsection{Clustering}
\label{ToC:TBL:Clustering}
Similar to the mixing layer, 
the clustering of the actuated boundary layer LES snapshots
is performed using a lossless POD compression.
Again, this compression dramatically reduces the computational load of clustering.
Here, we perform the POD and the clustering on all 38 test cases simultaneously.
Employing this enlarged set of POD modes yields richer, and thus more accurate, 
dynamical representation of the individual test cases
and allows for a direct comparison of different actuations  \citep{Ishar2019jfm}.
Concatenating all configurations results in $M=15873$ snapshots sampled at
$\Delta t = 0.94$ time units.

\begin{figure}[htb]
	\centering
	\def\svgwidth{1\linewidth}
	\includegraphics[width=0.9\linewidth]{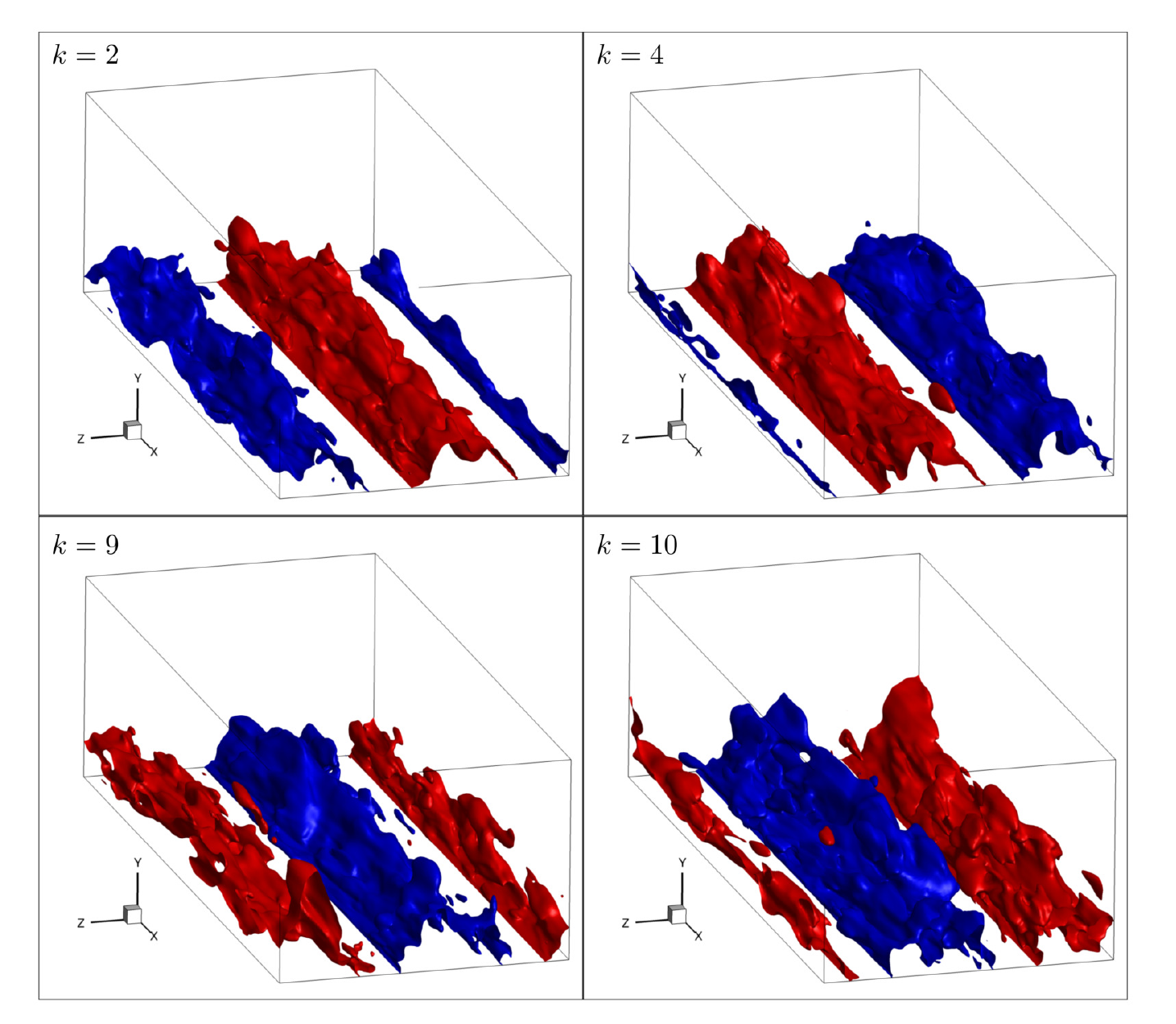}
	\caption{The cluster centroids $\bm{c}_{k}, k= 2, 4, 9, 10$ of the actuated boundary layer.
		The cluster numbers are denoted on the state-space figure \ref{fig:ProximityMap-BL}.
		The iso-surfaces correspond to constant wall-normal velocity of $V^+=\pm 0.04$. 
		Red and blue regions mark positive and negative values.}
	\label{fig:Centroids-BL}
\end{figure}
Following \citet{Ishar2019jfm},
the $M$ snapshots are clustered 
with the k-means++ algorithm into 50 centroids,
corresponding to $K=10$ centroids populated by the investigated actuation.
It is worth  noting that increasing $K$ significantly, say $K=100$, 
uncovers centroids with smaller length-scale
features associated with broadband turbulence of the boundary layer.
In this study, we purposely choose to focus on the main energy-containing dynamics
and thus limit the number of centroids to $K=10$.
Figure \ref{fig:Centroids-BL} presents four centroid distributions of the test case 
with  $\lambda^+=1000$, $T^+ =120$, and  $A^+ =40$.
As the figure shows, the centroids have similar spatial distributions
and are phase-shifted with respect to one another.
Such behaviour is consistent with a limit-cycle dynamics,
indicating partial lock-on of the boundary layer dynamics to the periodic surface actuation.	
This lock-on phenomenon is sometimes associated with aerodynamic gains or losses
depending on the targeted flow instability.
It is synonymous with synchronization, and has been repeatedly investigated
for drag reduction problems \citep{Barros2016,Taira2018,Herrmann2020}.
Similar to these studies, a lower actuation threshold with sufficient authority 
is required to synchronize the flow.

\begin{figure}[htb]
	\centering
	\includegraphics[width=0.7\linewidth]{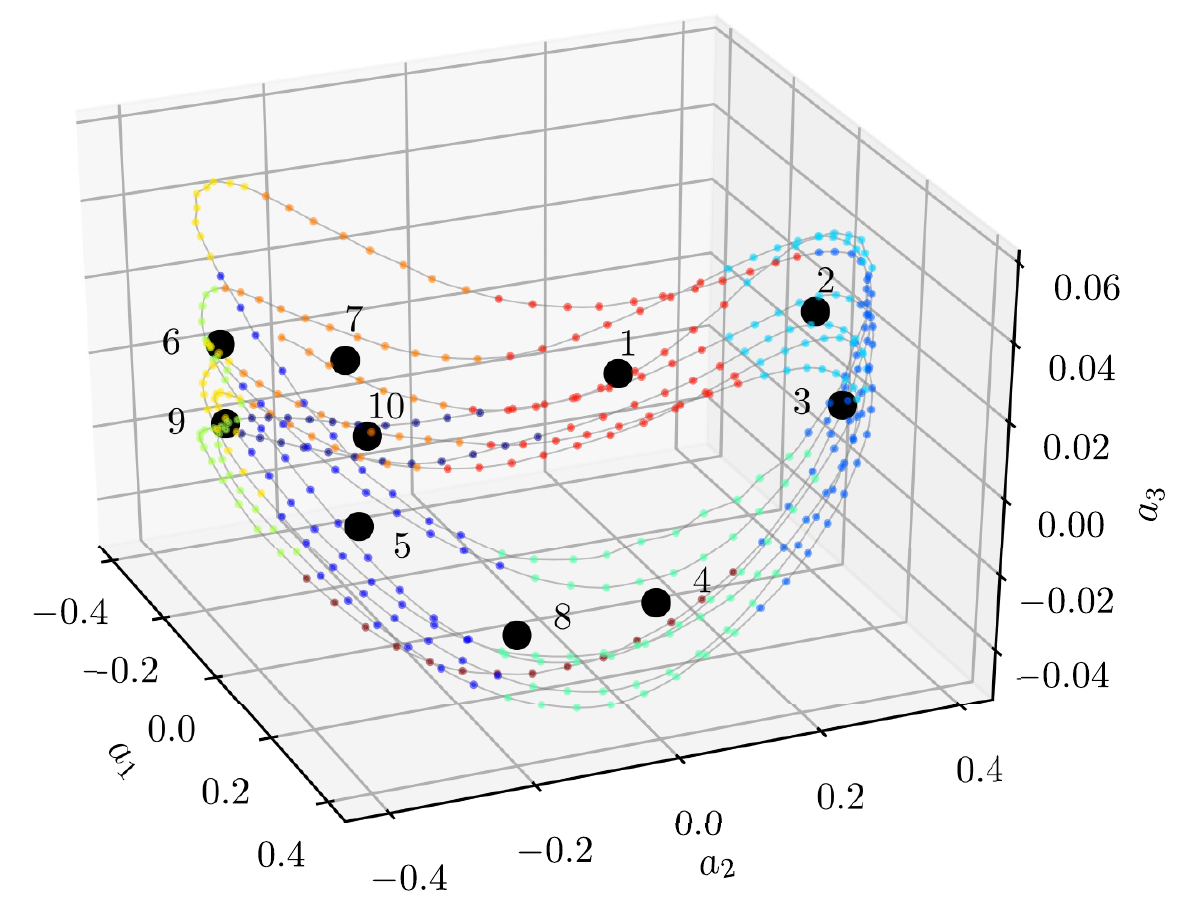}
	\caption{The state space is spanned by the first three 
	mode coefficients of the lossless proper orthogonal decomposition of the LES data. 
	The cluster centroids are displayed as black solid circles and the 
	snapshots are coloured according to their cluster affiliation.}
	\label{fig:ProximityMap-BL}
\end{figure}

The dynamics are well represented in the state space
illustrated in figure \ref{fig:ProximityMap-BL}, which is spanned by the first three POD mode coefficients.
The cluster centroids are displayed as black solid circles and their index is labelled.
The snapshots are coloured according to their cluster affiliation.
Similar to the shear layer, the dynamics of the actuated boundary layer appear to be driven by two physical phenomena:
a cyclic behaviour synchronized with the surface actuation, and a quasi-stochastic component 
that forces the limit cycle to experience cycle-to-cycle variations \citep{Cao2014ef}.
The latter phenomenon is associated with broadband turbulence of the boundary layer.

\subsection{Network model}
\label{ToC:TBL:CNN}

\begin{figure}
	\centering
	\sidecaption{subfig:a}
	\raisebox{-\height}{\def\svgwidth{0.445\linewidth}
	\includegraphics[width=0.44\linewidth]{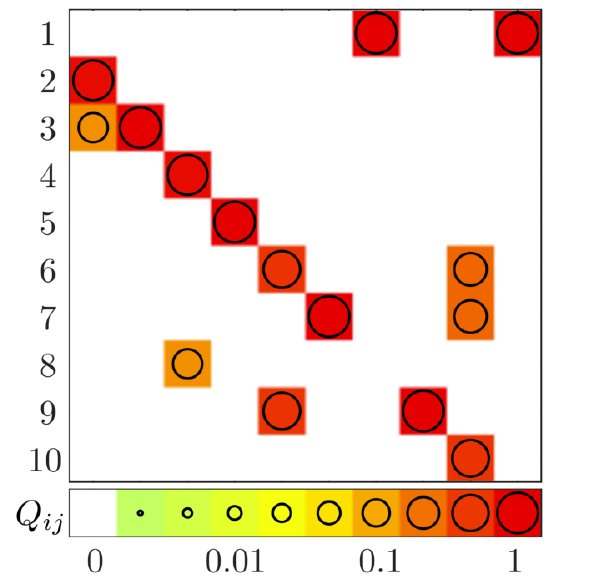}}
	\hfill
	\sidecaption{subfig:b}
	\raisebox{-\height}{\def\svgwidth{0.445\linewidth}
	\includegraphics[width=0.44\linewidth]{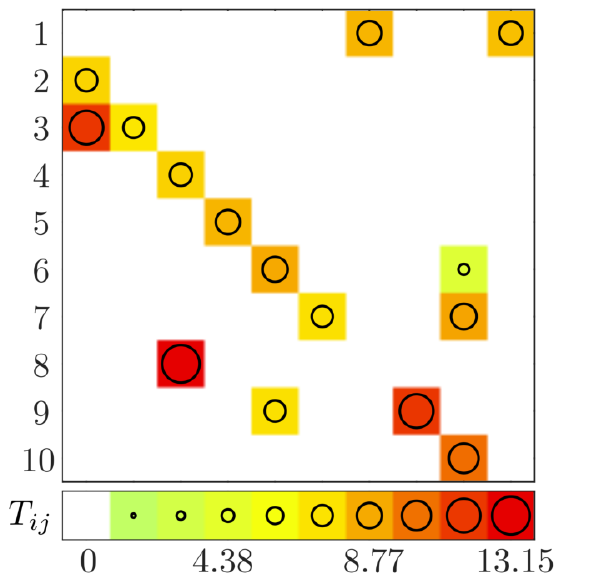}}
	\caption{ Dynamics of the cluster-based network model for the 
		actuated boundary layer.
		\subref{subfig:a} Direct transition matrix.  
		\subref{subfig:b} Averaged transition time. 
		The non-vanishing values of the matrix elements 
		are proportional to the circle radius 
		and can be inferred from the colour code from the bottom caption.}
	\label{fig:DirectTransitionMatrix-BL}
\end{figure}
The CNM is generated based on the direct transition matrix and the averaged transition time matrix,
which are illustrated in figure \ref{fig:DirectTransitionMatrix-BL}.
We reiterate the vanishing diagonal elements of both matrices, i.e., $Q_{ii}=T_{ii}=0$,
which is a result of enforcing non-trivial transitions.
The direct transition matrix (c.f. figure \ref{fig:DirectTransitionMatrix-BL}\subref{subfig:a}) 
shows both the dominant transition probability to subsequent centroids 
associated with the limit cycle behaviour,
and the wandering dynamics from the remaining transitions.
The transition time matrix between the centroids (c.f. figure \ref{fig:DirectTransitionMatrix-BL}\subref{subfig:b})
reflects the same behaviour, and exhibits a quasi-constant transition time for limit 
cycle ``flight times'' and diverse transition times for the wandering effect.

\begin{figure}
	\centering
	\def\svgwidth{0.8\linewidth}
	\includegraphics[width=0.8\linewidth]{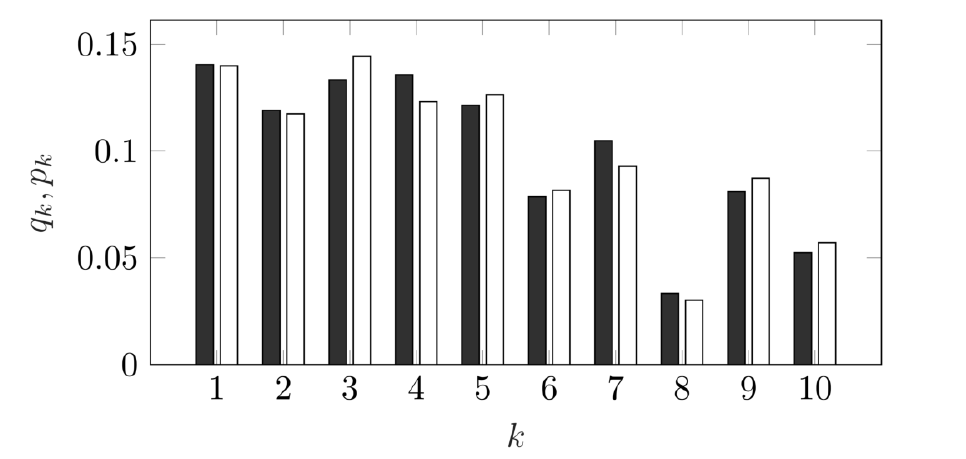}
	\caption{ Probability distribution of the actuated boundary layer
                from large eddy simulation (solid rectangle) and cluster-based network model (open rectangle).
                The CNM values are obtained from simulating  $l=106$ clusters transitions.}
	\label{fig:ProbCNM_Data_TBL}
\end{figure}

Figure \ref{fig:ProbCNM_Data_TBL} 
compares the probability distribution of LES data and the
model-based asymptotic vector $\bm{p}^\infty$.
Again, we choose cluster $k = 1$ as the initial condition for LES and for the CNM
and integrate over $l = 106$ transitions, which correspond to a similar time
range as that of the snapshots.
The agreement between the two distributions is good.

\begin{figure}
	\centering
	\sidecaption{subfig:a}
	\raisebox{-\height}{\def\svgwidth{0.3\linewidth}\includegraphics[width=0.44\linewidth]{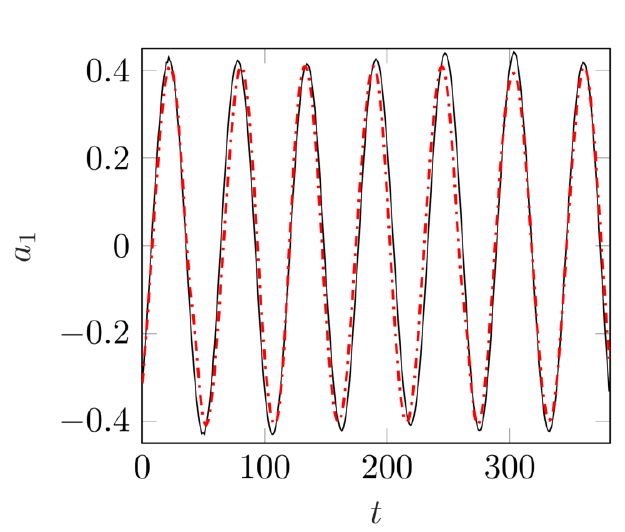}}%
	\hfill
	\sidecaption{subfig:b}
	\raisebox{-\height}{\def\svgwidth{0.3\linewidth}\includegraphics[width=0.44\linewidth]{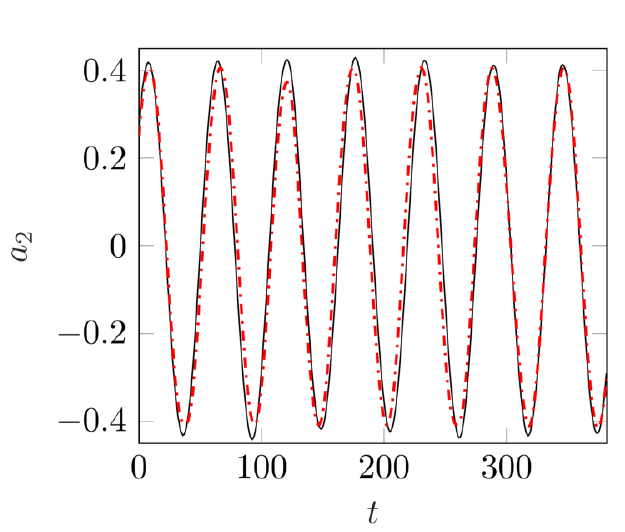}}
	\par
	\sidecaption{subfig:c}
	\raisebox{-\height}{\def\svgwidth{0.2\linewidth}\includegraphics[width=0.44\linewidth]{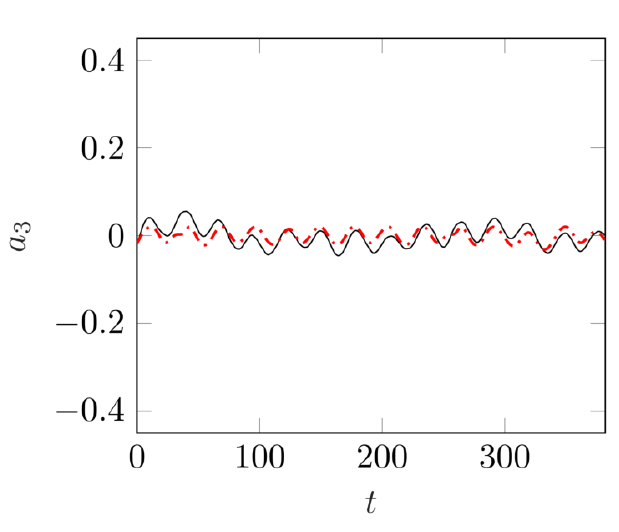}}%
	\hfill
	\sidecaption{subfig:d}
	\raisebox{-\height}{\def\svgwidth{0.2\linewidth}\includegraphics[width=0.44\linewidth]{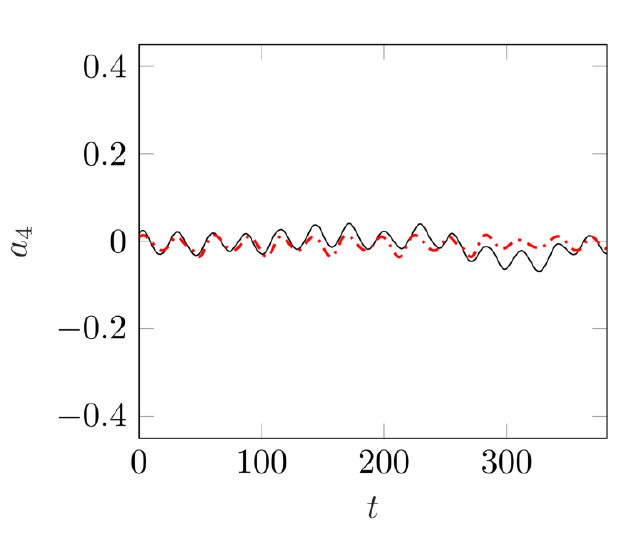}}%
	\caption{Evolution of mode amplitudes $a_{1}$--$a_{4}$ for the actuated boundary layer $t\in[0,400]$. 
		The curves correspond to LES (black solid line) and cluster-based network model (red dashed line).}
	\label{fig:ModeCoef-BL}
\end{figure}
The model performance is assessed against the reference LES results.
Figure \ref{fig:ModeCoef-BL} shows the evolution of the first four POD mode amplitudes (red dashed curve).
The dominance of the first two POD modes compared to the subsequent modes
is expected for the current quasi-synchronous actuated flow.
Similar to the previously-presented results, the temporal evolution is smoothed with a spline.
As the figure shows, CNM agrees very well with the amplitude and phase of the LES reference data 
over the entire approximately $400$ time units.

\begin{figure}[htb]
	\centering
        \def\svgwidth{0.3\linewidth}
	\includegraphics[width=1\linewidth]{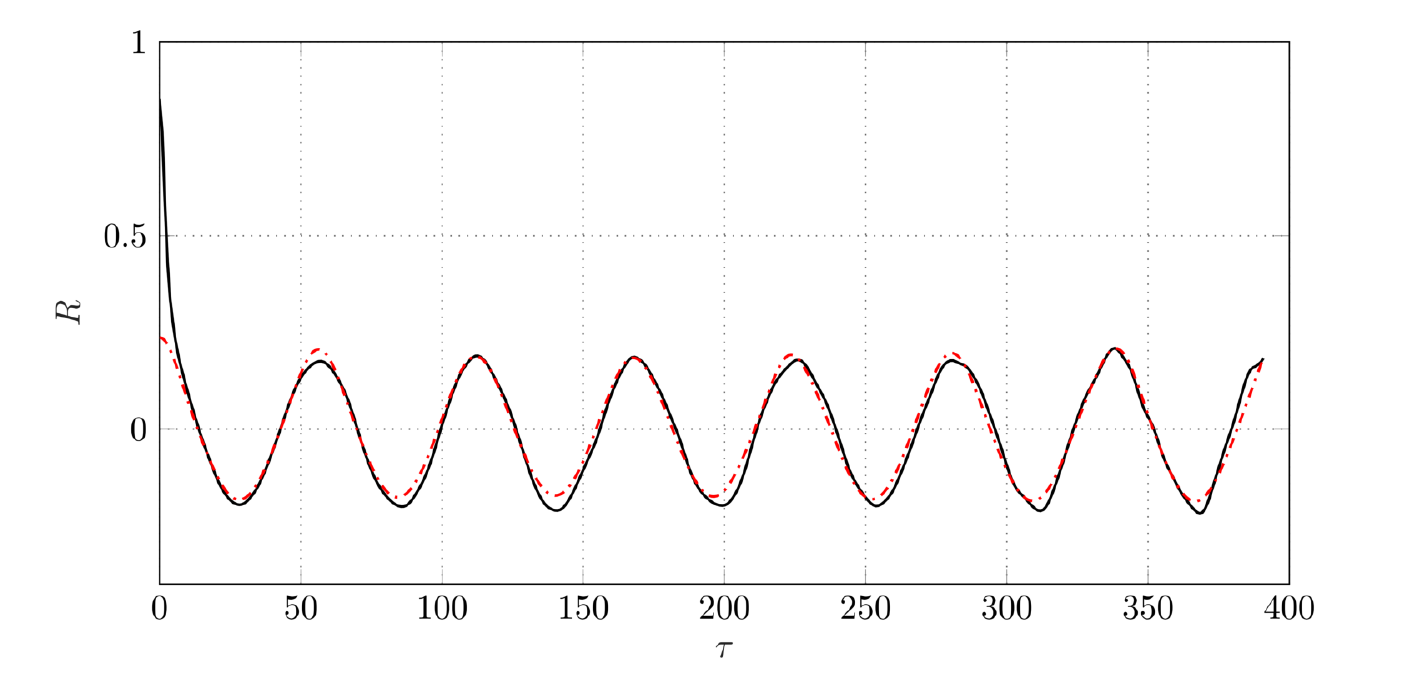}
	\caption{ 
	Autocorrelation function of the actuated turbulent boundary layer for LES (black solid line) and cluster-based network model  (red dashed line).}
	\label{fig:CorreFunction-BL}
\end{figure}
The agreement between the model and the reference data is further corroborated by
comparing the autocorrelation function.
Figure \ref{fig:CorreFunction-BL} displays the autocorrelation function of the CNM and the LES.
As with the mixing layer, 
the model-based fluctuation level at vanishing time delay 
is lower than the LES value
but becomes similar to oscillation level for an arbitrary larger time horizon.
This large representation error at $\tau =0$ 
relates to the unresolved inner-cluster variance.
Yet, the centroids adequately resolve the periodic flow response
of the flow to the periodic surface actuation.

\section{Conclusions}
\label{ToC:Conclusions}

In the present study, 
we propose a new data-driven methodology
for modelling nonlinear dynamical systems.
We trade compatibility with first principles,
like with a POD-based Galerkin model,
with the simplicity and robustness of the modelling.
Point of departure is the cluster-based Markov model \citep{Kaiser2014jfm}
for  time-resolved snapshot data.
The snapshots are coarse-grained into few representative centroids.
The temporal evolution of the state is conceptualized 
as straight a constant velocity movement
from one centroid to the next.
The average flight time and the transition probabilities
are inferred from the data.
Thus, the dynamics is modelled by a deterministic-stochastic network model
with the centroids as nodes, 
the straight trajectory segments as edges,
the transition time as parameters of the edges
and the transition probability characterizing the nodes.

The resulting cluster-based network model (CNM) has several desirable features:
(1) The methodology is simple and automatable.
(2) The off-line computational load is only slightly larger 
than a snapshot-based proper-orthogonal decomposition (POD). 
After the computation of the POD, 
the clustering and network model 
requires a tiny fraction of  the computational operation.
If the CNM is computed with original flow data without POD compression, 
the computational costs are orders of magnitudes larger
as elaborated in appendix \ref{ToC:Method:POD}.
(3) The CNM has the same recurrence properties as the original data:
If one cluster is visited multiple times in the data, 
it will also be a recurrence point of the CNM.
(4) Long-term integration will never lead to a divergence---unlike POD models.
(5) The framework is very flexible allowing, for instance, 
to incorporate multiple operating conditions.

The simplicity and robustness have a price.
On the kinematic side, the vanilla version of  CNM  does not have 
the possibility to extrapolate the data,
e.g., resolve oscillations at higher amplitudes not contained in the data.
On the dynamic side, we lose the relationship to first principles:
The network model is purely inferred from the snapshot data, without links to the Navier-Stokes equations.
In particular, cluster-based models are not natural frameworks 
for dynamic instabilities,
as the notion of exponential growth and nonlinear saturation is intimately tied to Galerkin flow expansions.
Subsequent generalizations need to overcome these restrictions.

Cluster-based network modelling (CNM) is applied to the Lorenz attractor.
A k-means++ algorithm yields 10 centroids from a long time-resolved solution.
4 centroids represent each ear of the attractor and 2 the switching area.
Despite the coarseness of the presentation, 
the temporal dynamics mimics well the oscillations in each ear
and the switching between both ears.
The agreement is mirrored by the similarity
between the autocorrelation functions of the simulation and the CNM. 
Statistically, the cluster population is predicted with acceptable accuracy.
The CNM dramatically outperforms the cluster-based Markov models (CMM) \citep{Kaiser2014jfm}
in terms of predicting the temporal evolution.
In contrast, CMM
is more accurate for the cluster population. 
The error source of the CNM can be traced back 
to the chosen simple model of transition times. 

Two demonstrations of CNM are performed with a laminar two-dimensional mixing layer
and with a periodically actuated turbulent boundary layer.
The mixing layer features Kelvin-Helmholtz (K-H)  vortices and occasional vortex pairing.
The cycle-to-cycle variations are clearly distilled by the centroids and
the proximity map shows the possible transitions.
The transition probabilities and times are quantified in the CNM model parameters.
The actuated turbulent boundary layer exhibits partial lock-on 
with a superimposed stochastic meandering.
For both applications, the snapshots are coarse-grained into 10  centroids.
For the mixing layer, one group of centroids can be associated with K-H vortices 
and a  second group to vortex pairing---similar to \citet{Kaiser2014jfm}.
In contrast, the centroid affiliations for the actuated turbulent boundary layer
are less categorizable.
The dominant periodic dynamics are superimposed with quasi-stochastic transitions
associated with broadband turbulence.
The CNM well resolves the temporal evolution of the main flow dynamics,
the fluctuation level, 
the autocorrelation function, and the cluster population.
A noteworthy observation relates to the autocorrelation function.
For vanishing time delay, this function displays the average representation
which is significant both for the mixing layer and wall turbulence.
Yet, the function is surprisingly well represented by the CNM after one characteristic period.
This behaviour corroborates that the dominant periodic dynamics 
is well resolved by the CNM with 10 centroids and the local interpolation between them.

CNM is found to have a distinct advantage over the departure point, CMM, 
namely the much longer prediction horizon
as evidenced by the autocorrelation function.
POD and DMD models may describe the same flow with a similar number of modes  \citep{Protas2015jfm}.
We emphasize that the construction of the CNM could be fully automated in a software package.
In contrast, data-driven nonlinear Galerkin models may 
be  designed as insightful least-order representations with interpretable modes.
Moreover, the Galerkin dynamics may reveal the interplay between linear and nonlinear terms,
as beautifully displayed in mean-field theory \citep{Stuart1971arfm}, 
self-consistent models \citep{Mantic2014prl}, 
resolvent operator approaches \citep{Gomez2016jfm},
finite-time thermodynamics \citep{Noack2008jnet} 
and criteria for boundedness \citep{Schlegel2015jfm}.
Yet, a functional model requires the careful choice
of flow data, potentially shift and other non-standard modes, 
subscale closure models and calibration techniques.
Thus, cluster-based and POD based models have different niche applications.

CNM opens a novel automatable avenue for nonlinear dynamical modelling.
Moreover, CNM  provides a framework for estimation and model-based control.
This extension 
is elaborated in appendix \ref{ToC:PODvsCNM} 
and complements  model-free cluster-based control
for open-loop actuation \citep{Kaiser2017tcfd} 
and for feedback laws \citep{Nair2019jfm}.
The authors actively pursue this direction.

\section*{Acknowledgements}

H.~L.\ appreciates the Graduate Student Research Innovation Project of Hunan Province (Grant No. CX2018B027). 
He gratefully acknowledges the support of the
China Scholarship Council (CSC)  (No. CSC201803170267) 
during his study in Technische Universit\"at Berlin 
and the excellent working conditions of the Hermann-F\"ottinger-Institute. 

D.~F., B.~R.~N., and R.~S. would like to thank the German science foundation (DFG)
grant number SE 2504/2-1 for supporting this work.
In addition, B.~R.~N. thanks the French National Research Agency 
(ANR-17-ASTR-0022 grant 'FlowCon'),
and the Bernd Noack Cybernetics Foundation
for additional support.
J.~T. acknowledges the funding 
from the National Natural Science Foundation of China (No. 91441121).
M.~M. acknowledges partial funding 
from the Polish Ministry of Science and Higher Education (MNiSW) under the Grant No.~05/54/DSPB/6492.

We are particularly indebted 
to Marian Albers and Wolfgang Schr\"oder 
for initiating our fruitful collaborative adventure on skin-friction reduction and 
for providing the employed LES data of the actuated boundary layer.
We have highly profited from stimulating discussions 
with Marian Albers,  
     Steven Brunton,
     Guy Yoslan Cornejo Maceda, 
     Nan Deng,  
     Arthur Ehlert, 
     Eurika Kaiser, 
     Matthew Lennie, 
     Qixin Lin, 
     Francois Lusseyran,  
     Christian Navid Nayeri, 
     Luc Pastur,  
     Christian Oliver Paschereit, 
     Wolfgang Schr\"oder, 
     and 
     Kunihiro Taira.
Last but not least, 
we thank the referees for important and insightful suggestions 
which have inspired new included investigations.

\section*{Declaration of interests}
The authors report no conflict of interest. 

\appendix

\section{Data compression for clustering}
\label{ToC:Method:POD}

Clustering is a computationally expensive process
based on a large number of area/vol\-ume integrals 
for the distance between snapshots and centroids. 
Let $M$ and $K$ be the amount of snapshots and clusters, respectively,
then a single k-means iteration requires the computation of $K \times M$ integrals. 
Let $I$ be the number of k-means iterations and 
$L$ be the number of repetitions 
then the total number of integrals is 
$L \times I \times K \times  M $.
Typical values are  $K\sim 10$, $I \sim 10 K$ and $L \sim 100$.

The computational load can be significantly reduced
by pre-processing the snapshot data with a lossless POD.
The most expensive step of a typical snapshot POD
is the computation of  the correlation matrix with $M \times (M+1)/2$ area/volume integrals.
Thus, 
the  integral for the distance between two velocity fields 
transforms into the Euclidean norm with $(M-1)$-dimensional vectors of POD mode amplitudes.
The computational saving reads
\begin{equation}
\label{Eqn:ComputationalSavings}
 \frac{M \times (M+1)/2}{L \times I \times K \times M} = \frac{M+1}{2 L \times I \times K}.
\end{equation}
With typical values, the savings are one or two orders of magnitudes.

For completeness and self-consistency, the snapshot POD algorithm is described.
POD is performed with the whole computational domain $\Omega$.
The inner product between two velocity fields $\bm{v}(\bm{x})$, $\bm{w}(\bm{x})$
in the square-integrable Hilbert space $ {\cal{L}}^2 (\Omega)$ reads
\begin{equation}
\left ( \bm{v} , \bm{w} \right )_{\Omega}  
= \int\limits_{\Omega} \! d \bf{x} \> \bm{v} ( \bm{x} ) \cdot \bm{w} ( \bm{x} ) 
\label{Eqn:InnerProduct}
\end{equation}
The corresponding norm is given by 
\begin{equation}
   \left \Vert \bm{v}  \right \Vert_{\Omega}  
=  \sqrt{ 
   \left (     \bm{v}, \bm{v} \right )_{\Omega} 
}.
\label{Eqn:Norm}
\end{equation}
The distance $D$ between two velocity fields is based on this norm,
\begin{equation}
   D( \bm{v}, \bm{w} )
=   \left \Vert \bm{v} - \bm{w} \right \Vert_\Omega .
\label{Eqn:Distance}
\end{equation}

The inner product \eqref{Eqn:InnerProduct}
uniquely defines the snapshot POD \citep[see, e.g.,][]{Holmes2012book}.
The $m$th snapshot is represented by 
\begin{equation}
\bm{u}^{m}(\bm{x}) :=\bm{u}_0 (\bm{x})+ 
\sum\limits_{i=1}^{M-1} a_i^m \bm{u}_{i}( \bm{x}),
\label{POD}
\end{equation}
where $\bm{u}_0$ denotes the mean flow,
$\bm{u}_i$ the $i$th POD mode and 
$a_i^m$ the POD mode amplitude corresponding to the $m$th snapshot. 
It may be noted that the maximal number of POD modes is $M-1$,
e.g., two snapshots define a one-dimensional line, not a plane.

Let $\bm{v}=\bm{u}_0 + \sum_{i=1}^{M-1} b_i \bm{u}_i$
and  $\bm{w}=\bm{u}_0 + \sum_{i=1}^{M-1} c_i \bm{u}_i$
be two velocity field representations, e.g., a snapshot and a centroid.
Then, their distance is given by
\begin{equation}
   D( \bm{v}, \bm{w} )_{\Omega}   =  \sqrt{\sum\limits_{i=1}^{M-1} \left (b_i -c_i \right )^2}.
\label{Eqn:Distance2}
\end{equation}
Evidently, 
\eqref{Eqn:Distance2} is much quicker to compute than \eqref{Eqn:Distance} 
assuming the typical case that the number of grid points is much larger than the number of snapshots.

\section{On the optimal number of clusters}
\label{Toc:Method:ModellingError}
We investigate the prediction error of a CNM 
with  direct transition matrix ${\sf{\bm{Q}}}$ 
and transition time matrix ${\sf{\bm{T}}}$ for $K$ clusters
from $M$ snapshot data.
The number of clusters $K$ significantly influences the prediction error of the CNM.
Coarse clustering (small $K$) means 
that the direct transition matrix $\boldsymbol{Q}$
can be inferred from a lot of transition data
and is hence relatively accurate.
Yet, the snapshots in each cluster have a large representation error.
In contrast, a  finely resolving clustering (large $K$) 
implies a more accurate representation of the true state. 
Yet,  the transition matrix is larger 
and the error of the estimated transition probability increases.
The extremes are $K=1$ cluster with large representation error
and $K=M$ with vanishing representation error, 
but large error of the transition matrix for new data.
We can expect a sweet spot with optimal prediction error
based on good representation error and an accurate estimate of the transition matrix.

In the following, we define the performance measure for the CNM. 
The starting point is the error between the model and true state 
$\delta \boldsymbol{u}(t)=\boldsymbol{u}^{\circ}(t)-\boldsymbol{u}^{\bullet}(t)$.
The modelling error for a specified number of clusters $K$ is defined as average error 
for all available snapshots with prediction horizon $\tau$ starting 
from the most accurate initial condition $\boldsymbol{u}^{\circ}(0)  \approx \boldsymbol{u}^{\bullet}(0)$.
The true initial state is taken from the snapshot data $\bm{u}^m$,
while the modelled initial state is the closest centroid $\bm{c}_{k^m}$.
The resulting error reads
\begin{equation}
C(\tau):=\overline{\left\|\boldsymbol{u}^{\circ}(t+\tau)-\boldsymbol{u}^{\bullet}(t+\tau)\right\|^{2}}_{\Omega}.
\label{Eqn:ModellingError}
\end{equation}
The overbar denotes the average over the prediction errors 
for all available snapshots $\boldsymbol{u}^{\bullet}(t)$
with data horizon until $t+ \tau$. 
$C(0)$ corresponds to the representation error where the 
true state $\boldsymbol{u}^{\bullet}(t)$ is estimated 
by the modeled state $\boldsymbol{u}^{\circ}(t)$ 
as accurately as possible. 
$C(\tau)$ is the prediction error after time $\tau$.

\begin{figure}[htb]
	\centering
	\def\svgwidth{0.3\linewidth}
	\includegraphics[width=\linewidth]{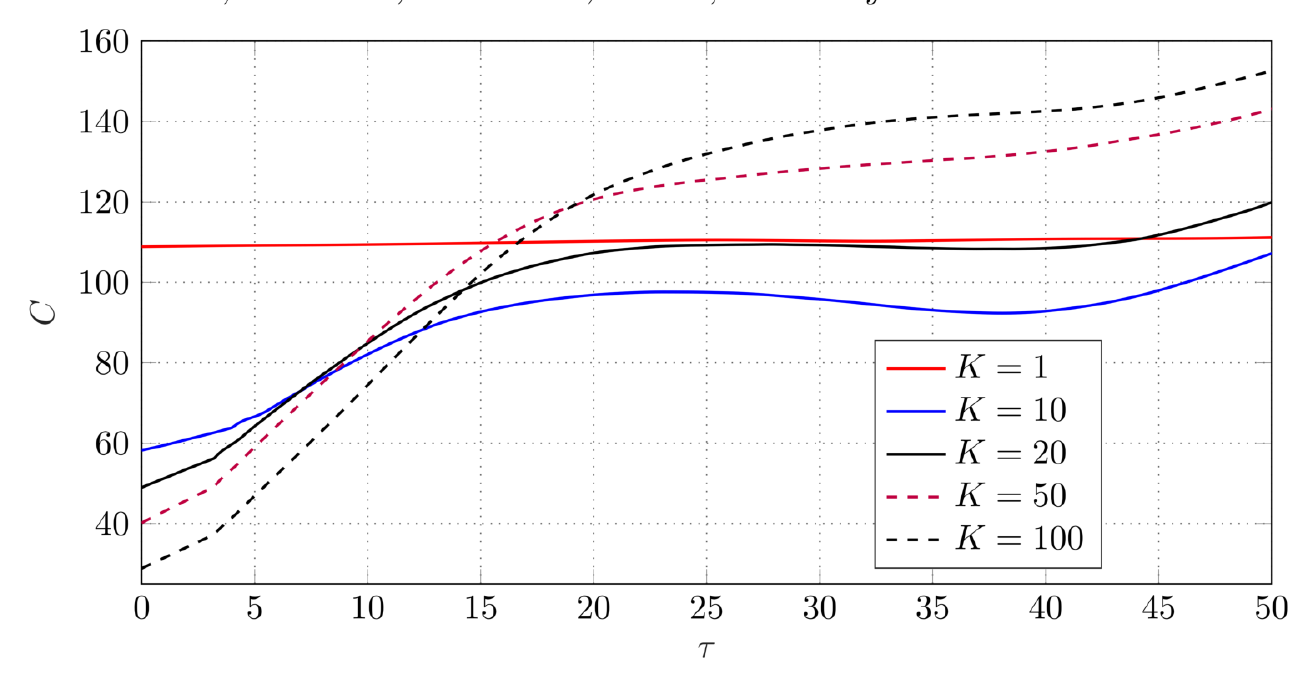}
	\caption{
	The prediction error $C(\tau)$ over time with selected number of clusters $K=1$, $10$, $20$, $50$, $100$.}
	\label{fig:ModellingError}
\end{figure}

Figure \ref{fig:ModellingError} illustrates the temporal evolution 
of prediction error for $\tau  \le 50$,
roughly corresponding to 5 Kelvin-Helmholtz shedding periods.
This error expectedly increases with
growing prediction horizon $\tau$ 
for all investigated numbers of clusters $K=10,20,50, 100$. 
There is no uniformly superior prediction error for any number of clusters. 
Small (large) $K$ correspond to  large (small) representation and prediction error for a small time horizon.
However, the more inaccurate transition matrix leads to larger prediction errors in the long run.
The CNM with $K=10$ leads to the smallest prediction error 
for $\tau \in [12.5,50]$
in comparison to all other investigated models.
Hence, we conclude that $K=10$ is a good choice for the cluster number
for a prediction horizon with one to several shedding periods.

We refrain from fine-tuning the optimal number of clusters
as this number is a function of the prediction horizon $\tau$.
For $\tau=0$, a CNM with $K=M$ reduces the representation error to zero, 
at least for the training data.
For $\tau=\infty$, the trivial CNM with $K=1$ yielding the mean flow 
outperforms all other models roughly by a factor 2.
 The average error between model and data 
 can easily be shown to be larger than the average distance of the data to the mean.
 Let us consider the data  $u^\bullet=\cos t$ 
 and model  $u^\circ = \cos 1.05 t$ with small frequency difference,
 i.e., increasing phase error.
 Then,  
 $\overline{ \left( u^\bullet-u^\circ \right)^2} = 1$ 
 but 
 $\overline{  \left( u^\circ  - 0 \right)^2 } = 1/2$.
 On average, $u^\circ$ stays closer to the mean $0$ than to another harmonics 
 which is occasionally out-of-phase. 
 
Finally, 
we remark that the number of clusters $K$ 
plays a similar role in cluster-based models
than the number of POD modes $N$ in Galerkin models.
Human interpretability is easier for a low-dimensional flow representation
while the accuracy increases with the model order.
For instance, for periodic dynamics, 
the phase resolution of each centroid  is approximately $360^\circ / K$.
However, there is a noticeable difference in robustness between CNM and POD models.
POD models tend to become more fragile with increasing state-space dimension,
as every new degree of freedom comes with many new coefficients
and potential error amplifiers.
In contrast, the robustness of cluster-based model does not suffer from increasing dimension.
A second difference relates to the modes.
Increasing the number of POD modes does not affect the lower-order modes by design.
In contrast, all centroids change as $K$ is just increased by $1$.
Similarly, all intervals of a one-dimensional finite-element discretization 
change as the number of elements increase by one.

\section{POD versus cluster-based network modelling}
\label{ToC:PODvsCNM}

POD models and CNM belong to the family 
of data-driven dynamic gray-box models 
which resolve the evolution of coherent structures.
Dynamic POD modelling was pioneered by \citet{Aubry1988jfm}
and has enjoyed over three decades of rapid development
on coherent structure descriptions, dynamical systems, estimation and control.
In contrast, networks have been recently introduced to reduced-order modelling
of fluid flows \citep{Nair2015jfm,Taira2016jfm}.
In this section, 
we compare POD models and CNM
with respect to kinematics (section \ref{ToC:Kinematics}),
dynamics (section  \ref{ToC:Dynamics}),
estimation (section  \ref{ToC:Estimation}) and
control (section  \ref{ToC:Control})---foreshadowing promising future opportunities of CNM.

\subsection{Kinematics}
\label{ToC:Kinematics}
Starting point of most data-driven gray-box models 
are  $M$ flow snapshots $\{ \bm{u} (\bm{x}) \}_{m=1}^M$
typically resolving first and second statistical moments.
Like the dynamic mode decomposition \citep{Rowley2009jfm,Schmid2010jfm},
the snapshots are assumed to resolve the coherent-structure evolution in time
so that the temporal dynamics can be identified.
POD expands the fluctuations around the mean flow $\bm{u}_0$
into a given number $N$ of orthonormal modes $\bm{u}_i$,
\begin{equation}
\label{Eqn:POD:GS}
\bm{u} (\bm{x}, t) = \bm{u}_0 (\bm{x}) + \sum\limits_{i=1}^N a_i(t) \> \bm{u}_i (\bm{x}) 
                                       +  \bm{\epsilon} (\bm{x},t).
\end{equation}
By design, this expansion minimizes the averaged representation error 
$\sum_{m=1}^M \Vert \bm{\epsilon} \Vert^2_{\Omega}/M$ with respect to all Galerkin expansions of $N$ modes.
The POD modes are linear combinations of the fluctuations $\bm{u}^m-\bm{u}_0$.

Clustering coarse-grains the snapshot data 
to a given number $K$ of centroids $\left \{ \bm{c}_k \right \}_{k=1}^K$.
Each snapshots with index $m$ belongs to the  closest centroid $k_m$.
The centroids are requested to minimize the averaged representation error 
$\Vert \bm{u}^m - \bm{c}_{k^m} \Vert^2$ .
Similar to POD modes, centroids are linear combinations of the snapshots.

The representation error of centroids 
can be further reduced by allowing for interpolations,
\begin{equation}
\label{Eqn:CM:GE}
\bm{u} (\bm{x}, t) = \sum\limits_{k=1}^K w_k(t) \> \bm{c}_k (\bm{x}),
\quad \sum\limits_{k=1}^K w_k(t) = 1,
\quad \forall k \colon w_k(t) \ge 0.
\end{equation}
In case of the Markov model, the weights are the evolving probabilities $w_k(t) = p_k(t)$
and make the expansion \eqref{Eqn:CM:GE} converge to the mean flow.
In case of the network model, the weights characterize `flights' with uniform velocity
between two centroids, say from $k$ to $j$, and typically re-visit all centroids in finite time.
The Markov model might be compared with unsteady Reynolds averaged Navier-Stokes (RANS) equations
converging to the mean flow
while the network model is reminiscent of the large-eddy simulations (LES). 

We emphasize that \eqref{Eqn:POD:GS} and \eqref{Eqn:CM:GE}
look similar but have quite different ranges of applications.
The POD expansion is based on the superposition of modes with arbitrary mode amplitudes $a_i$.
Neither the mean flow nor the POD modes are realizable states.
POD could be considered a data-driven analog of the Fourier expansion.
In contrast, 
the cluster-based expansion is only meant to describe a local interpolation for CNM.
The centroids are coarse-grained approximations of realizable states. 
The centroids may be conceptualized as collocation points for a finite-element inspired ansatz
and the associated Voronoi cells serve as finite elements. 
As a corollary, 
POD expansions can describe new states 
which are far from the snapshot database,
because the  mode amplitudes are not confined.
In contrast, cluster-based expansions are bound to stay close to the training data
by the non-negativity $w_k \ge 0$ and the normalization constraint $\sum_{k=1}^K  w_k = 1$.
By construction, global POD expansions have lower representation error
as cluster-based expansions with the same number of modes.
Some POD modes of simple dynamics may have a physical meaning 
as they resolve instability modes or harmonics.
Typically, however, POD modes comprise a mix of frequencies and are difficult to interpret.
In contrast, all centroids are human-interpretable 
coarse-grained flows which are representative for a certain state-space region.
Summarizing, the choice between POD and clustering 
strongly depends on the intended applications.

\subsection{Dynamics}
\label{ToC:Dynamics}
The temporal evolution of the incompressible viscous flow
can be derived from the Galerkin expansion 
and the Navier-Stokes equations \citep{Fletcher1984book}
for steady domains with stationary boundary conditions.
The resulting Galerkin system for the mode amplitude vector $\bm{a}= \left[ a_1, a_2, \ldots, a_N \right]^\mathrm{T}$ 
is the autonomous system
\begin{equation}
\label{Eqn:POD:GS}
\frac{d\bm{a}}{dt} = \bm{f} ( \bm{a} ).
\end{equation}
For turbulent flows, only a fraction of the fluctuation energy
is resolved by the POD modes and the effect of the remaining unresolved fluctuations
must be accounted for.
Myriad of subgrid turbulence models and calibration techniques have been proposed
and the identification of a robust realistic dynamical system constitutes a challenge.
Even the basic physical requirement of a globally bounded dynamics is often not met  \citep{Schlegel2015jfm}.

CNM might be conceptualized as flights between airports (centroids)
from a discrete network of routes with destination probabilities (transition matrix) and flight times (transition times).
In CNM, the chosen `destination' $j$ 
from `airport' $k$ at time $t^m$ 
from centroid $k$ at time $t^m$ 
 to centroid $j$ during $t^{m+1} = t^m + T_{jk}$ 
is described by
\begin{subequations}
\label{Eqn:CNM:Flight}
\begin{eqnarray}
j &=& \hbox{realization according to\ } Q_{jk}
\\ \bm{u} (\bm{x}, t) &=& w_k(t) \> \bm{c}_k( \bm{x} ) 
                         + w_j(t) \> \bm{c}_j( \bm{x} ),
\\  w_j(t) &=& (t-t^m)/T_{jk},
\quad
w_k(t) = 1-w_j(t).
\end{eqnarray}
\end{subequations}
At time $t^{m+1}$, a similar decision on the next destination is made, and so on.
The CNM \eqref{Eqn:CNM:Flight} describes a deterministic-stochastic dynamics
in contrast to the deterministic \eqref{Eqn:POD:GS}.

In contrast to POD models, 
the CNM \eqref{Eqn:CNM:Flight} contains no design parameter 
beyond the number of clusters and is fully automated.
Moreover, the dynamics is robust and cannot diverge, unlike POD models.
The price is the confinement to the neighbourhood of the training data.
Again, the decision in favor of the POD model or CNM 
strongly depends on the goal.
POD models may allow deeper dynamics insights.
CNM is much simpler and much more robust by design.

\subsection{Estimation}
\label{ToC:Estimation}
In most experiments, 
only few signals, denoted by the vector $\bm{s}(t)$, can be recorded.
Let 
$\bm{u}^m ( \bm{x}, t^m), m=1,\ldots, M$, be the snapshots associated
with the sensor readings $\bm{s}^m = \bm{s} (t^m)$.
The easiest realization of the estimator 
\begin{equation}
\label{Eqn:1NN:Estimator}
\hat{\bm{u}} (\bm{x},t) = \bm{G} \left ( \bm{x},  \bm{s}(t) \right )
\end{equation}
for sensor reading $\bm{s}$
is to find the closed sensor data $\bm{s}^m$ from the data base
and to take the corresponding snapshot $\bm{u}^m$ as an estimator.
This simplistic 1-nearest neighbour estimator 
can be refined in numerous ways.
An interpolation with $K$ data points can be performed
with a $K$-Nearest Neighbour approach \citep{Loiseau2018jfm}.
The sensor signals may be lifted to a feature space 
without dynamic false neighbours,
for instance with time-delay coordinates \citep{Loiseau2018jfm}.
Or the structure of $\bm{G}$ may be pre-assumed as in linear stochastic estimation.

The estimated flow field is canonically transcribed
 into POD mode amplitudes $\hat{\bm{a}}$
and permissible centroid weights 
$\hat{\bm{w}}=\left[ \hat{w}_1,\hat{w}_2,\ldots, \hat{w}_N \right]^\mathrm{T}$.
Summarizing, 
the estimation can easily be realized as add-ons
in POD models and CNM.
For completeness, 
we  mention the possibility of dynamic observers 
exploiting the dynamical system.

\subsection{Control}
\label{ToC:Control}
The POD models may be enriched with a forcing term.
In a simple case, 
like a volume force,
the forcing term is additive and linear 
in the actuation command $\bm{b} = \left [ b_1, b_2, \ldots, b_{N_b} 
\right]^T $ 
with the gain matrix $\bm{B}$,
\begin{equation}
\label{Eqn:POD:GS2}
\frac{d\bm{a}}{dt} = \bm{f} ( \bm{a} ) + \bm{B} \bm{b}.
\end{equation}
From here on, stabilizing control laws 
may be derived from linearizations  or other strategies \citep{Brunton2015amr}.

The control design for CNM is more complex.
The actuation command $\bm{b}$ affects 
the dynamics \eqref{Eqn:CNM:Flight}
via changed transition probabilities ${\sf{\bm{Q}}} (\bm{b}) $ 
and changed transition times ${\sf{\bm{T}}} (\bm{b})$.
\begin{equation}
\label{Eqn:CNM:Forcing}
{\sf{\bm{Q}}} = {\sf{\bm{Q}}}_0 + \sum\limits_{l=1}^{N_b} b_l \> {\sf{\bm{Q}}}_l, \quad
{\sf{\bm{T}}} = {\sf{\bm{T}}}_0 + \sum\limits_{l=1}^{N_b} b_l \> {\sf{\bm{T}}}_l.
\end{equation}
Here, the subscript `$0$' corresponds to the unforced state,
while the subscript `$l$' denotes changes caused by the actuation command $b_l$.
The matrices may be identified from actuated flow data.
After, 
the forced CNM \eqref{Eqn:CNM:Flight}\eqref{Eqn:CNM:Forcing} is identified, 
a regression solver can be employed to optimize the control law
with respect to a cost function.
Genetic programming has proven to be a powerful method for this method
in dozens of turbulence control experiments \citep{Noack2019springer}.

We remark that the stochastic-deterministic network dynamics
rules out `simple' control design based on local linearizations,
but requires the numerical solution of a non-convect nonlinear optimization problem.
Thus, the computational cost of this approach
 is significantly larger than the model-based linear control.
Yet, cluster-based network model may enable nonlinear infinite-horizon control
at a fraction of the computational cost of linear optimal control using the Navier-Stokes equations.
The authors actively pursue this novel avenue of cluster-based network control for turbulence.
\citet{Nair2019jfm} and \citet{Kaiser2017tcfd}  
present a model-free cluster-based control as a prelude to these efforts.

\clearpage
\bibliographystyle{jfm}

\end{document}